\documentclass[twocolumn,showpacs,aps,epsfig]{revtex4}

\usepackage{graphicx}
\usepackage{epstopdf}
\usepackage{latexsym}

\usepackage[center]{subfigure}

\begin{document}

 \newcommand{\bq}{\begin{equation}}
 \newcommand{\eq}{\end{equation}}
 \newcommand{\bqn}{\begin{eqnarray}}
 \newcommand{\eqn}{\end{eqnarray}}
 \newcommand{\nb}{\nonumber}
 \newcommand{\lb}{\label}
\newcommand{\PRL}{Phys. Rev. Lett.}
\newcommand{\PL}{Phys. Lett.}
\newcommand{\PR}{Phys. Rev.}
\newcommand{\CQG}{Class. Quantum Grav.}

\title{The hierarchy problem, radion mass, localization of gravity and
4D effective Newtonian potential in string theory on $S^{1}/Z_{2}$}
\author{  Anzhong Wang ${}^{1}$}   
\email{Anzhong_Wang@baylor.edu}  
\author{N.O. Santos ${}^{2,3}$} 
\email{N.O.Santos@qmul.ac.uk}
\affiliation{ ${}^{1}$ GCAP-CASPER, Department of Physics, Baylor University,
Waco, Texas 76798-7316\\
${}^{2}$ School of Mathematical Sciences, Queen Mary,
University of London, London E1 4NS, UK\\
${}^{3}$
Laborat\'orio Nacional de Computa\c{c}\~{a}o Cient\'{\i}fica, 
25651-070 Petr\'opolis RJ, Brazil}
 
\date{\today}

\begin{abstract}

In this paper, we present a systematical study of brane worlds of string theory
on $S^{1}/Z_{2}$. In particular,  starting with the toroidal compactification of 
the Neveu-Schwarz/Neveu-Schwarz sector  in (D+d) dimensions, we first obtain
an effective $D$-dimensional action, and then compactify one of the $(D-1)$ 
spatial dimensions by introducing two orbifold branes as its boundaries. We 
divide the whole set of the gravitational and matter field equations into two 
groups, one holds outside the two branes, and the other holds on them. By 
combining  the Gauss-Codacci and Lanczos equations, we write down explicitly the 
general gravitational field equations on each of the two branes, while using 
distribution theory we express the matter field equations on the branes in terms 
of the discontinuities of the first derivatives of the matter fields.  Afterwards,
we address three important issues: (i) the hierarchy problem; (ii) the radion 
mass; and (iii) the localization of gravity, the 4-dimensional Newtonian effective
potential and the Yukawa corrections due to the gravitational high-order Kaluza-Klein (KK)
modes.  The mechanism of solving the hierarchy problem is essentially the combination 
of the large extra dimension  and  warped factor mechanisms together with the 
tension coupling scenario. With very conservative arguments, we find that the
radion mass is of the order of $10^{-2}\; GeV$. The gravity is localized on the visible 
brane,  and the spectrum of the gravitational KK modes is discrete and can be of the order
of TeV. The corrections to the 4-dimensional Newtonian potential from the higher order 
of gravitational KK modes are exponentially  suppressed and can be safely neglected
in current experiments. In an appendix, we  also present  a systematical and pedagogical 
study of the Gauss-Codacci equations and Israel's junction conditions across a (D-1)-dimensional
hypersurface, which can be either spacelike or timelike.

\end{abstract}
\preprint{arXiv: xxxxxxxx}

\vspace{.7cm}


\maketitle

\vspace{1.cm}

\section{Introduction}

\renewcommand{\theequation}{1.\arabic{equation}}
\setcounter{equation}{0}

Superstring and M-theory all suggest that we may live in a world that 
has more than three spatial
dimensions.  Because only three of these are presently observable, one has to
explain why the others are hidden from detection.  One such explanation
is the so-called Kaluza-Klein (KK) compactification, according to which the
size of the extra dimensions is very small (often taken to be on the order
of the Planck length).  As a consequence, modes that have momentum in the
directions of the extra dimensions are excited at currently inaccessible
energies.

Recently, the braneworld scenarios \cite{ADD98,RS1} has dramatically changed
this point of view and, in the process, received a great deal of attention.
At present, there are a number of proposed models (See, for example,
\cite{branes} and references therein.). In particular,
Arkani-Hamed {\em et al} (ADD) \cite{ADD98} pointed out that the extra
dimensions need not necessarily be small and may even be on the scale of
millimeters.  This model assumes that Standard Model fields are confined
to a three (spatial) dimensional hypersurface (a 3-brane) living in a
larger dimensional bulk  while the gravitational field propagates
in the  bulk.  Additional fields may live only on the brane or
in the  bulk, provided that their current undetectability is 
consistent with experimental bounds. One of the most attractive features 
of this model is that it may potentially resolve the long standing 
hierarchy problem, namely the large difference in magnitudes
between the Planck and electroweak scales,
${M_{pl}}/{M_{EW}} \simeq 10^{16}$, 
where $M_{pl}$ denotes the four-dimensional Planck mass with $M_{pl} 
\sim 10^{16} \; TeV$, and $M_{EW}$  the electroweak scale with
$M_{EW} \sim  TeV$. 

Considering a N-dimensional spacetime and assuming that the extra dimensions 
are homogeneous and finite, we find
\bqn
\label{1.1}
  S^{(N)}_{g} &\sim& -{M^{N-2}}  \int{dx^{4} dz^{n}\sqrt{- g^{(N)}}
R^{(N)}} \nb\\
 &=&  -{M^{N-2}V_{n}}\int{d^{4}x  \sqrt{- g^{(4)}}
R^{(4)}}\nb\\
&\simeq& -{M^{2}_{pl} }\int{d^{4}x  \sqrt{- g^{(4)}}
R^{(4)}},
\eqn
where  $V_{n}$ denotes the volume of extra dimensions, $n \equiv N-4$, 
and $M$  the N-dimensional fundamental Planck mass, which is related to
$M_{pl}$ by  
\bq
\lb{1.2}
M = \left({M^{2}_{pl}}/{V_{n}}\right)^{1/(2+n)}.
\eq
Clearly, for any given   extra dimensions, if 
$V_{n}$  is large enough, $M$ can be as low as the electroweak scale,
$M \simeq M_{EW} \simeq TeV$. Therefore, if we consider  $M$ as the fundamental 
scale and $M_{pl}$ the deduced one, we can see that the hierarchy between
the two scales is exactly due to the dilution of the spacetime in high
dimensions, whereby the hierarchy problem is resolved.
Table top experiments show that Newtonian gravity is valid at least 
down to the size $R \sim 44 \; \mu{m}$ \cite{Hoyle04}. From the above 
we can see that for $n \ge 2$ the N-dimensional Planck mass $M$ can
be lowered down to the electroweak scale  from the four-dimensional 
Planck scale.

In a different model, Randall and Sundrum (RS1) \cite{RS1}
showed that if the self-gravity of the brane is included, gravitational
effects can be localized near the Planck brane at low energy and the 4D Newtonian 
gravity is reproduced on this brane. 
In this model, the extra dimensions are not homogeneous,
but warped. One of the most attractive features of the 
model is that it will soon be fully explored by LHC \cite{DHR00}. 
In the RS1 scenario, the mechanism to solve the hierarchy
problem is completely different \cite{RS1}. Instead of using large
dimensions, RS used the warped factor, for which the mass $m_{0}$
measured on the invisible (Planck) brane is related to the mass 
$m$ measured on the visible 
(TeV) brane by $m = e^{-ky_{c}}m_{0}$, where
$e^{-ky_{c}}$ is the warped factor. 
Clearly, by properly choosing the distance $y_{c}$ 
between the two branes, one can lower $m$ to the order of $TeV$, 
even $m_{0}$ is of the order of $M_{pl}$. 
It should be noted that the five-dimensional Planck mass $M_{5}$
in the RS1 scenario is still in the same order of $M_{pl}$. In fact,  
the 5-dimensional action $S^{5)}_{g}$ can be written as
\bqn
\lb{1.3}
S^{(5)}_{g} &\sim& -{M^{3}}\int{dx^{4} d\phi\sqrt{- g^{(5)}}
R^{(5)}}\nb\\
 &=& -{M^{3}}\int^{\pi}_{-\pi}{r_{c}e^{-2kr_{c}|\phi|}d\phi}
\int{d^{4}x \sqrt{- g^{(4)}}
R^{(4)}} \nb\\
&\simeq& -{M^{2}_{pl} }\int{d^{4}x  \sqrt{- g^{(4)}}
R^{(4)}},
\eqn
where now we have
\bq
\lb{1.4}
 M^{2}_{pl} = {M^{3}}{k^{-1}}\left(1 - e^{-2ky_{c}}\right) \simeq M^{2}_{5},
 \eq
 for $k \simeq M_{5}$ and $ky_{c} \simeq 35$.
 
Another long-standing problem  is the cosmological constant problem: Its theoretical 
expectation values  
from quantum field theory exceed its observational limits by $120$ orders of 
magnitude  \cite{wen}. Even if such high energies are suppressed by supersymmetry, 
the electroweak corrections  are still $56$ orders higher. 
This problem was further sharpened by recent observations of supernova (SN) Ia, 
which  reveal the revolutionary  discovery that our universe has lately been in 
its accelerated expansion phase  \cite{agr98}. Cross checks from the cosmic microwave 
background radiation  and large scale structure all confirm it \cite{Obs}. 
In Einstein's theory of gravity, such an expansion can be achieved by a tiny positive 
cosmological constant, which is well consistent with all observations carried out 
so far \cite{SCH07}. Because of this remarkable result, a large number of ambitious 
projects have been aimed to distinguish the cosmological constant from dynamical 
dark energy models \cite{DETF}. 
Since the problem is intimately related to quantum gravity, its solution 
is expected to come from quantum gravity, too. At the present, string/M-Theory is our 
best bet for a consistent  quantum theory of gravity, so it is  reasonable to ask 
what string/M-Theory has to say about the cosmological constant. 

In the string landscape \cite{Susk}, it is expected there are many different vacua 
with different local cosmological constants \cite{BP00}. Using the anthropic principle, 
one may select the low energy vacuum in which we can exist. However, many theorists 
still hope to explain the problem without invoking the existence of ourselves. 
In addition, to have a late time accelerating universe from string/M-Theory, 
Townsend and Wohlfarth \cite{townsend} invoked a time-dependent compactification  
of pure gravity in higher dimensions with hyperbolic internal space to circumvent 
Gibbons' non-go theorem \cite{gibbons}. Their exact solution   exhibits 
a short period of acceleration. The solution is the zero-flux limit of spacelike branes 
\cite{ohta}. If non-zero flux or forms are turned on,  a transient acceleration exists 
for both compact internal hyperbolic and flat spaces \cite{wohlfarth}. Other 
accelerating solutions  by compactifying more complicated time-dependent internal 
spaces can be found  in \cite{string}.

Recently,  we studied brane cosmology in the framework of  both string theory \cite{WS07,WSVW08} 
and the Horava-Witten (HW) heterotic M Theory  \cite{GWW07,WGW08}  on $S^{1}/Z_{2}$. From a
pure numerology point of view, we found that the   4D  effective cosmological constant  can be cast in the form,  
\bq
\lb{1.5}
\rho_{\Lambda}
= \frac{\Lambda_{4}}{8\pi G_{4}}
= 3\left(\frac{R}{l_{pl}}\right)^{\alpha_{R}}\left(\frac{M}{M_{pl}}\right)^{\alpha_{M}}
M_{pl}^{4},
\eq
where $R$ denotes the typical size of the extra dimensions, $M$ is the energy scale of string or M theory,
 and $(\alpha_{R}, \alpha_{M}) = (10, 16)$ for string
theory and $(\alpha_{R}, \alpha_{M}) = (12, 18)$ for the HW heterotic M Theory. In both 
cases, it can be shown that for $R \simeq 10^{-22} \; m$ and $M_{10} 
\simeq 1\; TeV$, we obtain $\rho_{\Lambda} \sim \rho_{\Lambda, ob} \simeq 10^{-47}\; 
GeV^{4}$.  

When orbifold branes are concerned, a critical ingredient is the radion stability. 
Using the mechanism of Goldberger and Wise \cite{GW99}, we showed that the radion is
stable.  Such studies were also generalized to the HW heterotic M Theory
\cite{HW96}, and found that, among other things, the radion is stable and has a mass of order
of $10^{-2} \; GeV$ \cite{WGW08}.

In this paper, we shall give a systematical study of brane worlds of string theory
on $S^{1}/Z_{2}$. Similar studies in 5-dimensional spacetimes have been
carried out in the framework of both string theory \cite{WSVW08} and M Theory
\cite{WGW08}. However, to have this paper as much independent as possible, it is difficult
to  avoid repeating some of our previous materials, although we would  try
our best to keep it to its minimum. The rest of the paper is organized as follows: 
In Sec. II, we consider the toroidal compactification of the Neveu-Schwarz/Neveu-Schwarz 
(NS-NS) sector  in (D+d) dimensions, and obtain an effective $D$-dimensional action.
Then, we compactify one of the $(D-1)$ spatial dimensions by introducing two 
orbifold branes as the boundaries along this compactified dimension. 
In Sec. III, we divide the whole set of the gravitational and matter field equations 
into two groups, one holds outside the two branes, and the other holds on each of them. 
Combining  the Gauss-Codacci and Lanczos equations, we write down explicitly the general 
gravitational field equations on the branes, while using distribution theory we
are able to express the matter field equations on the branes in terms of the 
discontinuities of the first derivatives of the matter fields. 
In Sec. IV, we study the hierarchy problem, while in Sec. V, we consider the radion mass 
by using  the Goldberger-Wise mechanism \cite{GW99}. In Sec. VI we study the localization 
of gravity, the 4-dimensional effective potential and high order Yukawa corrections. In   
Sec. VII,  we present our main conclusions with some discussing
remarks. We also include an appendix, in which we present a systematical and pedagogical
study of the Gauss-Codacci equations and Israel's junction conditions across a surface,
where the metric coefficients are only continuous, i.e., $C^{0}$, in higher dimensional  
spacetimes. To keep such a treatment as general as possible, the surface can be either
spacelike or timelike.

Before turning to the next section, we would like to note that  in 4-dimensional spacetimes
there exists Weinberg's no-go theorem for the adjustment of the cosmological constant 
\cite{wen}. However, in higher dimensional spacetimes, the 4-dimensional vacuum energy on the 
brane does not necessarily give rise to an effective 4-dimensional cosmological constant. 
Instead, it may only curve the bulk, while leaving the brane still flat \cite{CEG01}, 
whereby Weinberg's no-go theorem is evaded. It was exactly in this vein, 
the cosmological constant problem was studied in the framework of brane worlds in 
5-dimensional spacetimes \cite{5CC} and 6-dimensional supergravity \cite{6CC}. However, 
it was soon found that in the 5-dimensional case hidden fine-tunings are required \cite{For00}.
In the 6-dimensional case such fine-tunings may not be needed, but it is still not 
clear whether loop corrections can be as small as  expected \cite{Burg07}.

\section{The Model}
\renewcommand{\theequation}{2.\arabic{equation}}
\setcounter{equation}{0}

In this section, we  first consider the toroidal compactification of the 
NS-NS sector  in (D+d) dimensions, and obtain an effective $D$-dimensional action. Then, 
we  compactify one of the $(D-1)$ spatial dimensions
by introducing two orbifold branes as the boundaries along this compactified dimension.

\subsection{Toroidal Compactification of the  NS-NS sector }

Let us consider  the  NS-NS sector  in (D+d) dimensions, 
$\hat{M}_{D+d} = M_{D}\times {\cal{T}}_{d}$, where  ${\cal{T}}_{d}$
is a d-dimensional torus. Topologically, it is the Cartesian product of d circles, 
${\cal{T}}_{d} = S^{1}\times S^{1}\times ... \times S^{1}$. Then, the action takes the form 
\cite{LWC00,BW06,MG07},
\bqn
\lb{2.1}
\hat{S}_{D+d} &=& - \frac{1}{2\kappa^{2}_{D+d}}
\int{d^{D+d}x\sqrt{\left|\hat{g}_{D+d}\right|}  e^{-\hat{\Phi}} \left\{
{\hat{R}}_{D+d}[\hat{g}]\right.}\nb\\
& & \left. + \hat{g}^{AB}\left(\hat{\nabla}_{A}\hat{\Phi}\right)
\left(\hat{\nabla}_{B}\hat{\Phi}\right) - \frac{1}{12}{\hat{H}}^{2}\right\},
\eqn
where $\hat{\nabla}_{A}$ denotes the covariant derivative with  respect to $\hat{g}^{AB}$
with  $A, B = 0, 1, ..., D+d-1$, and $\hat{\Phi}$ is the dilaton field. The NS three-form 
field $\hat{H}_{ABC}$ is defined as
\bqn
\lb{2.1a}
\hat{H}_{ABC} &=& 3 \partial_{[A}\hat{B}_{BC]}\nb\\
&=& \partial_{A}\hat{B}_{BC} + \partial_{B}\hat{B}_{CA}
+ \partial_{C}\hat{B}_{AB},
\eqn
where  the square brackets imply total antisymmetrization over all indices, and
\bq
\lb{2.1b}
 \hat{B}_{CD} = - \hat{B}_{DC}, \;\;\;
 \partial_{A}\hat{B}_{CD}
 \equiv \frac{\partial\hat{B}_{CD}}{\partial{x^{A}}}. 
 \eq
The constant $\kappa^{2}_{D+d}$ denotes the gravitational coupling constant, defined as
\bq
\lb{2.2}
\kappa^{2}_{D+d} = 8\pi G_{D+d} = \frac{1}{\left(M_{D+d}\right)^{D+d-2}},
\eq
where $G_{D+d}$ and $M_{D+d}$ denote, respectively,  the ($D+d$)-dimensional Newtonian constant
and  Planck mass. 

In this paper we consider the $(D+d)$-dimensional spacetimes 
described by the metric,
\bqn
\lb{2.3}
d{\hat{s}}^{2}_{D+d} &=& \hat{g}_{AB} dx^{A}dx^{B} \nb\\
&=&
   \tilde{g}_{ab}\left(x\right) dx^{a}dx^{b}  +   h_{ij}\left(x\right)dz^{i} dz^{j},
\eqn
where  $\tilde{g}_{ab}(x)$ is the metric on $M_{D}$, parametrized by the coordinates $x^{a}$
with  $a,b, c = 0, 1, ..., D-1$, and $h_{ij}(x)$ is the metric on the compact space ${\cal{T}}_{d}$
with  periodic coordinates $z^{i}$, where $i, j = D, D+1, ..., D+d-1$. 

We assume that all the matter fields, similar to the metric coefficients, are functions of
$x^{a}$ only, 
\bq
\lb{2.4}
\hat{\Phi} = \hat{\Phi}\left(x^{a}\right), \; \;\;\;
\hat{B}_{CD} = \hat{B}_{CD}\left(x^{a}\right).
\eq
This implies that the compact space  ${\cal{T}}_{d}$ is Ricci flat,
\bq
\lb{2.5}
R_{d}[h] = 0,
\eq
and that
\bq
\lb{2.5a}
 \hat{H}_{ijk} =  3 \partial_{[i}\hat{B}_{jk]} = 0.
\eq
For the sake of simplicity, we also assume that  the flux $\hat{B}$ is block diagonal, 
\bq
\lb{2.5c}
\left(\hat{B}_{MN}\right) = \left(\matrix{\tilde{B}_{ab}(x) & 0\cr
0 & B_{ij}(x)\cr}\right).
\eq
Then, it can be shown that 
\bqn
\lb{2.5b}
\hat{H}_{abc} &=& \tilde{H}_{abc} = 3 \partial_{[a}\tilde{B}_{bc]}, \nb\\
\hat{H}_{aij} &=& \tilde{\nabla}_{a}{B}_{ij}, \;\;\;
\hat{H}_{abi} = 0,
\eqn
where  $\tilde{\nabla}_{a}$ denotes the covariant derivative with  respect to 
$\tilde{g}^{ab}$.

On the other hand, we also have  
\bqn
\lb{2.6}
{\hat{R}}_{D+d}[\hat{g}] &=& R_{D}[\tilde{g}] 
    + \frac{1}{4}\left(\tilde{\nabla}_{a}h^{ij}\right)\left(\tilde{\nabla}^{a}h_{ij}\right)\nb\\
    & &
    +  \tilde{\nabla}_{a}\left(\ln\sqrt{|h|}\right)\tilde{\nabla}^{a}\left(\ln\sqrt{|h|}\right)
    \nb\\
  & & - \frac{2}{\sqrt{|h|}}\tilde{g}^{ab}\tilde{\nabla}_{a}\tilde{\nabla}_{b}\left(\sqrt{|h|}\right).
\eqn 
Inserting Eq.(\ref{2.6}) into Eq.(\ref{2.1}) and then integrating the internal part, we obtain
the effective $D-$dimensional action,
\bqn
\lb{2.7}
S_{D} &=& - \frac{1}{2\kappa^{2}_{D}}
\int{d^{D}x\sqrt{\left|\tilde{g}_{D}\right|} e^{-\tilde{\phi}}
\left\{\tilde{R}_{D}[\tilde{g}] + \left(\tilde{\nabla}_{a}\tilde{\phi}\right)
\left(\tilde{\nabla}^{a}\tilde{\phi}\right)\right.}\nb\\
& & + \frac{1}{4} \left(\tilde{\nabla}_{a}h^{ij}\right)\left(\tilde{\nabla}^{a}h_{ij}\right)
- \frac{1}{12}\tilde{H}_{abc}\tilde{H}^{abc}\nb\\
& & \left. - \frac{1}{4} h^{ik}h^{jl}\left(\tilde{\nabla}_{a}B_{ij}\right)
\left(\tilde{\nabla}^{a}B_{kl}\right)\right\},
\eqn
where
\bqn
\lb{2.8a}
\tilde{\phi} &=& \hat{\Phi} - \frac{1}{2}\ln\left|h\right|,\\
\lb{2.8b}
\kappa^{2}_{D} &\equiv& \frac{\kappa^{2}_{D+d}}{V_{0}},
\eqn
with  the $d-$dimensional internal volume given by
\bq
\lb{2.9}
{\cal{V}}_{d}\left(x^{a}\right) \equiv \int{d^{d}z \sqrt{|h|}} =  |h|^{1/2} V_{0}.
\eq
Action (\ref{2.7}) is usually referred to as that written in the string frame.

To go to the Einstein frame, we make the following conformal transformations,
\bqn
\lb{2.10}
g_{ab} &=& \Omega^{2}\tilde{g}_{ab},\nb\\
\Omega^{2} &=& \exp\left(-\frac{2}{D-2}\tilde{\phi}\right),\nb\\
\phi &=& \sqrt{\frac{2}{D-2}} \; \tilde{\phi}.
\eqn
Then, the action (\ref{2.7}) takes the form
\bqn
\lb{2.11}
S_{D}^{(E)} &=& - \frac{1}{2\kappa^{2}_{D}}
\int{d^{D}x\sqrt{\left|{g}_{D}\right|}  
\left\{{R}_{D}[{g}] - \frac{1}{2}\left(\nabla\phi\right)^{2}\right.}\nb\\
& &  + \frac{1}{4} \left({\nabla}_{a}h^{ij}\right)\left({\nabla}^{a}h_{ij}\right)\nb\\
& & - \frac{1}{12} e^{- \sqrt{\frac{8}{D-2}}\phi}H_{abc}H^{abc}\nb\\
& & \left. - \frac{1}{4} h^{ik}h^{jl}\left({\nabla}_{a}B_{ij}\right)
\left({\nabla}^{a}B_{kl}\right)\right\},
\eqn
where $\nabla_{a}$ denotes the covariant derivative with  respect to $g_{ab}$. It should be
noted that, since the definition of the three-form $\hat{H}_{ABC}$ given by (\ref{2.1a}) is 
independent of the metric, it is conformally invariant. In particular, we have
\bq
\lb{2.11aa}
H_{abc} = \tilde{H}_{abc},\;\;
B_{ab} = \tilde{B}_{ab}.
\eq
However, we do have
\bqn
\lb{2.11a}
& & {H}^{abc} = g^{ad}g^{be}g^{cf}H_{def} = \Omega^{-6}\tilde{H}^{abc},\nb\\
& & H_{abc}H^{abc} = \Omega^{-6}\tilde{H}_{abc}\tilde{H}^{abc}. 
\eqn

Considering the addition of a potential term  \cite{BW06},  in the string frame we have
\bq
\lb{2.12}
\hat{S}^{m}_{D+d} = - \int{d^{D+d}x \sqrt{\left|\hat{g}_{D+d}\right|} V_{D+d}^{s}}.
\eq
  Then, after the dimensional reduction
 we find
\bq
\lb{2.13}
{S}_{D, m} = - V_{0} \int{d^{D}x \sqrt{\left|\tilde{g}_{D}\right|} \; {|h|}^{1/2} V_{D+d}^{s}},
\eq
where
\bq
\lb{2.14}
\tilde{g}_{D} = \exp\left(\sqrt{\frac{2D^{2}}{D-2}}\; \phi\right)\; g_{D}.
\eq
Changed to the Einstein frame, the action (\ref{2.13}) becomes
\bq
\lb{2.13a}
{S}_{D, m}^{(E)} = - \frac{1}{2\kappa^{2}_{D}} \int{d^{D}x \sqrt{\left|{g}_{D}\right|}   V_{D}},
\eq
where
\bq
\lb{2.14a}
V_{D} \equiv 2\kappa^{2}_{D}V_{0}V_{D+d}^{s} 
\exp\left(\frac{D}{\sqrt{2(D-2)}}\;\phi\right)\; |h|^{1/2}.
\eq

If we further assume that
\bqn
\lb{2.15}
h_{ij} &=& - \exp\left(\sqrt{\frac{2}{d}}\; \psi\right) \delta_{ij},\nb\\ 
h^{ij} &=& - \exp\left(-\sqrt{\frac{2}{d}}\; \psi\right) \delta^{ij},
\eqn
we find that
\bqn
\lb{2.16}
S_{D}^{(E)} + S_{D, m}^{(E)} &=& - \frac{1}{2\kappa^{2}_{D}}
\int{d^{D}x\sqrt{\left|{g}_{D}\right|}  
\left\{{R}_{D}[{g}] \right.}\nb\\
& & - \frac{1}{2}\left[\left(\nabla\phi\right)^{2} + \left(\nabla\psi\right)^{2}
- 2V_{D}\right]\nb\\
& &  - \frac{1}{4} e^{- \sqrt{\frac{8}{d}}\; \psi}
  \left({\nabla}_{a}B_{ij}\right)
\left({\nabla}^{a}B^{ij}\right)\nb\\
& & \left. - \frac{1}{12} e^{- \sqrt{\frac{8}{D-2}}\; \phi}H_{abc}H^{abc}\right\},
\eqn
where $B^{ij} \equiv \delta^{ik}\delta^{jl}B_{kl}$, and the effective $D-$dimensional potential 
(\ref{2.14}) now is given by
\bq
\lb{2.17}
V_{D} \equiv 2\kappa^{2}_{D}V_{0} V_{D+d}^{s} \exp\left(\frac{D}{\sqrt{2(D-2)}}\;\phi 
 + \sqrt{\frac{d}{2}}\;\psi\right).
\eq

\subsection{$S^{1}/Z_{2}$ Compactification of the D-Dimensional Sector}

We shall compactify  one of the $(D-1)$ spatial dimensions by putting two orbifold branes
as its boundaries. The brane actions are taken as,
\bqn
\lb{3.1}
S^{(E, I)}_{D-1, m} &=& -  \int_{M^{(I)}_{D-1}}{\sqrt{\left|g^{(I)}_{D-1}\right|}
\left(\epsilon_{I}V^{(I)}_{D-1}(\phi, \psi) + g^{(I)}_{\kappa}\right)} \nb\\
& &  \times d^{D-1}\xi_{(I)}\nb\\
& & + \int_{M^{(I)}_{D-1}}{d^{D-1}\xi_{(I)}\sqrt{\left|g^{(I)}_{D-1}\right|}}\nb\\
& & \times {\cal{L}}^{(I)}_{D-1,m}\left(\phi, \psi, B,\chi\right),
\eqn
where $I, J = 1, 2,\; V^{(I)}_{D-1}(\phi, \psi)$  denotes the potential of the scalar 
fields $\phi$ and $\psi$ on the branes, and $\xi_{(I)}^{\mu}$'s are the intrinsic coordinates of the 
branes with  $\mu, \nu = 0, 1, 2, ..., D-2$, and  $\epsilon_{1} = - \epsilon_{2} = 1$.
$\chi$  denotes collectively the matter fields, and $ g^{(I)}_{\kappa}$ denotes the tension of
the I-th brane. As to be shown below,  it is directly related to the $(D-1)$-dimensional 
Newtonian constant $G_{D-1}^{(I)}$ \cite{Cline99}. 
The two branes are localized on the surfaces,
\bq
\lb{3.3d}
\Phi_{I}\left(x^{a}\right)  = 0,
\eq 
or equivalently
\bq
\lb{3.3db}
x^{a} = x^{a}\left(\xi^{\mu}_{(I)}\right).
\eq 
$g^{(I)}_{D-1}$ denotes the determinant of the reduced metric  $g_{\mu\nu}^{(I)}$  of the I-th  brane, 
defined as
\bq
\lb{3.3c}
g_{\mu\nu}^{(I)} \equiv \left. g_{ab} e^{(I)a}_{(\mu)} e^{(I)b}_{(\nu)}\right|_{M^{(I)}_{D-1}}, 
\eq
where
\bq
\lb{3.3dd}
e^{(I)\; a}_{(\mu)} \equiv \frac{\partial x^{a}}{\partial \xi^{\mu}_{(I)}}. 
\eq

Then,  the total action is given by,
\bq
\lb{3.2}
S^{(E)}_{total} = S_{D}^{(E)} + S_{D, m}^{(E)} + \sum_{I=1}^{2}{S^{(E, I)}_{D-1, m}}.
\eq

\section{Field Equations Both Outside and on the Orbifold Branes}
\renewcommand{\theequation}{3.\arabic{equation}}
\setcounter{equation}{0}

Variation of the total action (\ref{3.2}) with respect to   the metric ${g}_{ab}$ yields 
the  field equations,
\bqn
\lb{3.3}
G^{(D)}_{ab} &=& \kappa^{2}_{D}T^{(D)}_{ab} + \kappa^{2}_{D} 
\sum^{2}_{I=1}{{\cal{T}}^{(I)}_{\mu\nu} e^{(I, \; \mu)}_{a}e^{(I, \; \nu)}_{b}}\nb\\
& &  \times \sqrt{\left|\frac{g^{(I)}_{D-1}}{g_{D}}\right|} \;
\delta\left(\Phi_{I}\right),
\eqn
where  $\delta(x)$ denotes the Dirac delta function, normalized in the sense of \cite{LMW01},
and the energy-momentum tensors $T^{(D)}_{ab}$ and ${\cal{T}}^{(I)}_{\mu\nu}$ are defined as,
\bqn
\lb{3.3a}
\kappa^{2}_{D}T^{(D)}_{ab} &\equiv & \frac{1}{2}
\left[\left(\nabla_{a}\phi\right)\left(\nabla_{b}\phi\right)
 + \left(\nabla_{a}\psi\right)\left(\nabla_{b}\psi\right)\right.\nb\\
& & + \frac{1}{2}e^{-\sqrt{\frac{8}{d}}\; \psi}  
\left(\nabla_{a}B^{ij}\right)\left(\nabla_{b}B_{ij}\right)\nb\\
& & \left. + \frac{1}{2}e^{\sqrt{\frac{8}{D-2}}\; \phi} H_{acd}H_{b}^{\;\; cd}\right]\nb\\
& & - \frac{1}{4}g_{ab}
\left[\left(\nabla\phi\right)^{2} + 
\left(\nabla\psi\right)^{2}  - 2V_{D}\right. \nb\\
& &   \frac{1}{2}e^{-\sqrt{\frac{8}{d}}\; \psi}  
\left(\nabla_{c}B^{ij}\right)\left(\nabla^{c}B_{ij}\right)\nb\\
& & \left. + \frac{1}{6}e^{\sqrt{\frac{8}{D-2}}\; \phi} H_{cde}H^{cde}\right],\\
\lb{3.3aa}
{\cal{T}}^{(I)}_{\mu\nu} &\equiv& \tau^{(I)}_{\mu\nu} 
+  \left(g^{(I)}_{\kappa} + \tau^{(I)}_{(\phi,\psi)} \right)g_{\mu\nu}^{(I)},\nb\\
\tau^{(I)}_{\mu\nu} &\equiv& 2 \frac{\delta{\cal{L}}^{(I)}_{D-1,m}}
{\delta{g^{(I)\; \mu\nu}}}
-  g^{(I)}_{\mu\nu}{\cal{L}}^{(I)}_{D-1,m},
\eqn
where   
\bqn
\lb{3.3b}
& & \tau^{(I)}_{(\phi,\psi)} \equiv  \epsilon_{I} V^{(I)}_{D-1}(\phi,\psi), \nb\\
& & e^{(I, \mu)}_{a} \equiv  \left. g^{(I)\; \mu\nu}e^{(I)\; b}_{(\nu)}  
g_{ab}\right|_{M^{(I)}_{4}},\nb\\
& & g^{(I)\; \mu\nu} g^{(I)}_{\lambda\nu} = \delta^{\mu}_{\lambda}.
\eqn

Variation of the total action (\ref{3.2}), respectively, with  respect to 
$\phi, \; \psi, \; B_{ij}$ and $B_{ab}$, yields the following equations of the 
matter fields,
\bqn
\lb{3.ea}
\Box\phi &=& - \frac{\partial{V_{D}}}{\partial{\phi}} 
- \frac{1}{12}\sqrt{\frac{8}{D-2}}e^{-\sqrt{\frac{8}{D-2}}\; \phi}H_{abc}H^{abc}\nb\\
& & - 2\kappa_{D}^{2}\sum^{2}_{i=1}{\left(\epsilon_{I} \frac{\partial{V_{D-1}^{(I)}}}
{\partial{\phi}} + \sigma^{(I)}_{\phi}\right)}\nb\\
& & \times\sqrt{\left|\frac{g^{(I)}_{D-1}}{g_{D}}\right|} \; \delta\left(\Phi_{I}\right),\\
\lb{3.eb}
\Box\psi &=& - \frac{\partial{V_{D}}}{\partial{\psi}} 
- \sqrt{\frac{1}{2d}} e^{-\sqrt{\frac{8}{d}}\; \psi}\left(\nabla_{a}B^{ij}\right)
\left(\nabla^{a}B_{ij}\right)\nb\\
& & - 2\kappa_{D}^{2}\sum^{2}_{i=1}{\left(\epsilon_{I} \frac{\partial{V_{D-1}^{(I)}}}
{\partial{\psi}} + \sigma^{(I)}_{\psi}\right)}\nb\\
& & \times\sqrt{\left|\frac{g^{(I)}_{D-1}}{g_{D}}\right|} \; \delta\left(\Phi_{I}\right),\\
\lb{3.ec}
\Box B_{ij} &=& \sqrt{\frac{8}{d}}\left(\nabla_{a}\psi\right)\left(\nabla^{a}B_{ij}\right)\nb\\
& & -  \sum^{2}_{i=1}{\sigma^{(I)}_{ij}
\sqrt{\left|\frac{g^{(I)}_{D-1}}{g_{D}}\right|} \; \delta\left(\Phi_{I}\right)},\\
\lb{3.ed}
\nabla^{c}H_{cab} &=& \sqrt{\frac{8}{D-2}}\; H_{cab}\nabla^{c}\phi\nb\\
& & -  \sum^{2}_{i=1}{\sigma^{(I)}_{ab}
\sqrt{\left|\frac{g^{(I)}_{D-1}}{g_{D}}\right|} \; \delta\left(\Phi_{I}\right)},
\eqn
where $\Box \equiv g^{ab}\nabla_{a}\nabla_{b}$ and
\bqn
\lb{3.ee}
\sigma_{\phi}^{(I)} &\equiv& -  \frac{\delta{\cal{L}}^{(I)}_{D-1,m}}{\delta\phi},\nb\\
\sigma_{\psi}^{(I)} &\equiv& -  \frac{\delta{\cal{L}}^{(I)}_{D-1,m}}{\delta\psi},\nb\\
\sigma^{(I)}_{ij} &\equiv& - 4\kappa^{2}_{D} e^{\sqrt{\frac{8}{d}}\; \psi}
\frac{\delta{\cal{L}}^{(I)}_{D-1,m}}{\delta{B^{ij}}},\nb\\
\sigma^{(I)}_{ab} &\equiv& - 4\kappa^{2}_{D} e^{\sqrt{\frac{8}{D-2}}\; \phi}
\frac{\delta{\cal{L}}^{(I)}_{D-1,m}}{\delta{B^{ab}}}.
\eqn
 
Eq.(\ref{3.3}) and Eqs.(\ref{3.ea})-(\ref{3.ed}) consist of the complete set of the gravitational 
and matter field equations.
To solve these equations, it is found very convenient to separate them  into two groups: (a) one is 
defined outside the two orbifold branes; and (b) the other is defined on the two branes. 

\subsection{Field Equations Outside the Two Branes}

To write down the equations outside the two orbifold branes is straightforward, and they are simply  the 
D-dimensional Einstein field equations (\ref{3.3}), and the matter field equations
Eqs.(\ref{3.ea})-(\ref{3.ed})  without the delta function parts,
\bqn
\lb{3.ef}
G^{(D)}_{ab} &=& \kappa^{2}_{D}T^{(D)}_{ab},\\
\lb{3.eg}
\Box\phi &=& - \frac{\partial{V_{D}}}{\partial{\phi}} \nb\\
& &- \frac{1}{12}\sqrt{\frac{8}{D-2}} e^{-\sqrt{\frac{8}{D-2}}\; \phi}H^{2},\\
\lb{3.eh}
\Box\psi &=& - \frac{\partial{V_{D}}}{\partial{\psi}}  \nb\\
& &
- \sqrt{\frac{1}{2d}} e^{-\sqrt{\frac{8}{d}}\; \psi}\left(\nabla_{a} B^{ij}\right)^{2},\\
\lb{3.ei}
\Box B_{ij} &=& \sqrt{\frac{8}{d}}\left(\nabla_{a}\psi\right)\left(\nabla^{a}B_{ij}\right),\\
\lb{3.ej}
\nabla^{c}H_{cab} &=& \sqrt{\frac{8}{D-2}}\; H_{cab}\nabla^{c}\phi,  
\eqn 
where $T^{(D)}_{ab}$ is given by Eq.(\ref{3.3a}). Therefore, in the rest of this section, we shall 
concentrate ourselves on the derivation of the field equations on the branes. 

\subsection{Field Equations on the Two Branes}

To write down the field  equations on the two orbifold branes, one can follow two different 
approaches: (1)  First express the delta function parts in the left-hand sides of
Eqs.(\ref{3.3}) and (\ref{3.ea})-(\ref{3.ed}) in terms of the discontinuities of the first 
derivatives of the metric coefficients and matter fields, and then equal the corresponding
delta function parts  in the right-hand sides of these equations, as shown systematically in 
\cite{WCS06}. (2) The second approach is to use the Gauss-Codacci and Lanczos equations to write 
down the $(D-1)$-dimensional gravitational field equations on the branes  \cite{SMS}. It should be noted 
that these two  approaches are  equivalent and complementary one to the other. In this paper, 
we shall follow the second approach to write down the gravitational field equations on the 
two branes, and the first approach to write the matter field equations on the two branes.

\subsubsection{Gravitational Field Equations on the Two Branes}

For  timelike branes, their normal vectors are spacelike. Then, setting $\epsilon(n) = -1$
in (\ref{C.6}) we obtain, 
 \bq
 \lb{3.12}
 G^{(D-1)}_{\mu\nu} = {\cal{G}}^{(D)}_{\mu\nu} + E^{(D)}_{\mu\nu}
 + {\cal{F}}^{(D-1)}_{\mu\nu},
 \eq
 with
 \bqn
 \lb{3.13}
 {\cal{G}}_{\mu\nu}^{(D)} &\equiv&  \frac{D-3}{(D-2)}
\left\{G_{ab}^{(D)}e^{a}_{(\mu)} e^{b}_{(\nu)} \right.\nb\\
& & \left.
- \left[G_{ab}n^{a}n^{b} + \frac{1}{D-1} G^{(D)}\right]g_{\mu\nu}\right\}, \nb\\
E^{(D)}_{\mu\nu} &\equiv& C_{abcd}^{(D)}n^{a}e^{b}_{(\mu)}n^{c}e^{d}_{(\nu)},\nb\\
{\cal{F}}^{(D-1)}_{\mu\nu} &\equiv&  
 K_{\mu\lambda}K^{\lambda}_{\nu} - KK_{\mu\nu} \nb\\
& & - \frac{1}{2}g_{\mu\nu}\left(K_{\alpha\beta}K^{\alpha\beta} 
    - K^{2}\right),
\eqn
where $n^{a}$ denotes the normal vector to the brane, $G^{(D)}
\equiv g^{ ab} G^{(D)}_{ab}$, and $C_{abcd}^{(D)}$ the Weyl tensor. 
The extrinsic curvature $K_{\mu\nu}$ is defined as
\bq
\lb{3.13a}
K_{\mu\nu} \equiv e^{a}_{(\mu)}e^{b}_{(\nu)}\nabla_{a}n_{b}.
\eq
A crucial step of this approach is the Lanczos equations \cite{Lan22},
\bq
\lb{3.4}
\left[K_{\mu\nu}^{(I)}\right]^{-} - g_{\mu\nu}^{(I)} \left[K^{(I)}\right]^{-} 
= - \kappa^{2}_{D}{\cal{T}}_{\mu\nu} ^{(I)},
\eq
where 
\bqn
\lb{3.5}
\left[K_{\mu\nu}^{(I)}\right]^{-} &\equiv& {\rm lim}_{\Phi_{I} \rightarrow 0^{+}}
K^{(I)\; +}_{\mu\nu} - {\rm lim}_{\Phi_{I} \rightarrow 0^{-}}
K^{(I)\; -}_{\mu\nu},\nb\\
\left[K^{(I)}\right]^{-} &\equiv& g^{(I)\; \mu\nu}\left[K_{\mu\nu}^{(I)}\right]^{-}.
\eqn

Assuming that the branes have $Z_{2}$ symmetry, we can express the intrinsic
curvatures $K^{(I)}_{\mu\nu}$ in terms of the effective energy-momentum tensor
${\cal{T}}_{\mu\nu} ^{(I)}$ through the Lanczos equations (\ref{3.4}). Setting
\bq
\lb{3.14}
 {\cal{S}}^{(I)}_{\mu\nu} = \tau^{(I)}_{\mu\nu} + g_{\kappa}^{(I)}g^{(I)}_{\mu\nu},
 \eq
 where $g_{\kappa}^{(I)}$ is a constant, which will be uniquely determined by
 the $(D+d)$- and $(D-1)$-dimensional gravitational coupling constants $\kappa_{D+d}$
 and $\kappa_{D-1}$ via Eqs.(\ref{2.8b}) and (\ref{3.17}), we find that
 \bq
\lb{3.14aa}
 {\cal{T}}^{(I)}_{\mu\nu} = \tau^{(I)}_{\mu\nu} + \left(g_{\kappa}^{(I)}
 + \tau^{(I)}_{(\phi, \psi)}\right) g^{(I)}_{\mu\nu}.
 \eq
 Then,
 $ G^{(D-1)}_{\mu\nu}$ given by Eq.(\ref{3.12}) can be cast in the form
 [cf. Eq.(\ref {D.18})],
 \bqn
 \lb{3.15}
 G^{(D-1)}_{\mu\nu} &=& {\cal{G}}^{(D)}_{\mu\nu} + E^{(D)}_{\mu\nu}
 + {\cal{E}}_{\mu\nu}^{(D-1)} + \kappa^{4}_{D}\pi_{\mu\nu}\nb\\
 & & + \kappa^{2}_{D-1}\tau_{\mu\nu} + \Lambda_{D-1} g_{\mu\nu},
 \eqn
 where
 \bqn
\lb{3.16}
\pi_{\mu\nu} &\equiv& \frac{1}{4}\left\{\tau_{\mu\lambda}\tau^{\lambda}_{\nu}
-  \frac{1}{D-2}\tau \tau_{\mu\nu}\right.\nb\\
& & \left. 
 - \frac{1}{2}g_{\mu\nu}\left(\tau^{\alpha\beta} \tau_{\alpha\beta}
 - \frac{1}{D-2}\tau^{2}\right)\right\},\nb\\
 {\cal{E}}_{\mu\nu}^{(D-1)} &\equiv& \frac{\kappa^{4}_{D}(D-3)}{4(D-2)}\tau_{(\phi,\psi)}\nb\\
 & & \times \left[\tau_{\mu\nu}  
   + \left(g_{\kappa} + \frac{1}{2}\tau_{(\phi,\psi)}\right)g_{\mu\nu}\right],
\eqn
and 
\bqn
\lb{3.17}
\kappa^{2}_{D-1} &=& \frac{D-3}{4(D-2)}g_{\kappa}\kappa^{4}_{D},\nb\\
\Lambda_{D-1}  &=&  \frac{D-3}{8(D-2)}g_{\kappa}^{2}\kappa^{4}_{D}.
\eqn
 For a perfect fluid,
\bq
\lb{3.18}
\tau_{\mu\nu} = \left(\rho + p\right)u_{\mu}u_{\nu} - p g_{\mu\nu},
\eq
where $u_{\mu}$ is the four-velocity of the fluid, we find that 
\bqn
\lb{3.19}
\pi_{\mu\nu} &=&  \frac{D-3}{4(D-2)}\rho\nb\\
& & \times \left[\left(\rho + p\right)u_{\mu}u_{\nu} 
- \left(p + \frac{1}{2}\rho\right)g_{\mu\nu}\right].
\eqn
Note that in writing Eqs.(\ref{3.15})-(\ref{3.19}), without causing any confusion, we had 
dropped the super indices $(I)$.  

It should be noted that in writing down Eqs.(\ref{3.15})-(\ref{3.17}) we implicitly assumed that 
only the brane tension has contribution to the (D-1)-dimensional Newtonian constant. 
However,  it was argued that when the scalar field does not vanish, it also contributes to
it \cite{BB03}. While this seems  reasonable, considering the fact that the tension
$g_{\kappa}$ has the same contribution to $G_{D-1}$, as one can see from Eqs.(\ref{3.14aa}),
there are several disadvantages for
such an inclusion: (i) The resulted Newtonian constant usually depends not only
on time but also on space, $G_{D-1} = G_{D-1}(\phi(t, x^{i}))$, which is highly constrained
experimentally \cite{CZ01}. (ii) It is model-dependent. Different potentials of the scalar field
on the brane will give different  $G_{D-1}$. (iii) It is not unique, even after the potential
is fixed. In fact, one can always redefine the energy-momentum tensor $ \tau^{(I)}_{\mu\nu}$
so that $ \tau^{(I)}_{\mu\nu} = \tilde{ \tau}^{(I)}_{\mu\nu} + \lambda^{(I)}g^{(I)}_{\mu\nu}$, where
the $\lambda^{(I)}$ term in Eq.(\ref{3.14aa}) takes the same form as  $g_{\kappa}^{(I)}$ and
 $ \tau^{(I)}_{(\phi, \psi)}$ do. Then, since both $\lambda^{(I)}$ and $ \tau^{(I)}_{(\phi, \psi)}$ are
 due to matter fields on the branes, there is no reason to assume that $\lambda^{(I)}$ has no
 contribution to $G_{D-1}$ but  $ \tau^{(I)}_{(\phi, \psi)}$ does. Therefore, in this paper,
 we shall take the point of view of \cite{Cline99}, and assume that only brane tension 
 couples with $G_{D-1}$. With such an assumption, it can be seen that $G_{D-1}$ is 
 uniquely defined once the brane tension is specified.

\subsubsection{Matter Field Equations on the Two Branes}

On the other hand, the I-th brane, localized on the surface $\Phi_{I}(x) = 0$, divides the
spacetime into two regions, one with $\Phi_{I}(x) > 0$ and the other with $\Phi_{I}(x) < 0$
[cf. Fig. 1]. Since the field equations are the second-order differential equations, the
matter fields have to be at least continuous across this surface, although in general their 
first-order directives are not. Introducing the Heaviside function, defined as
\bq
\lb{3.21a}
H\left(x\right) = \left\{\matrix{1, & x > 0,\cr
0, & x < 0,\cr} \right.
\eq
in the neighborhood of $\Phi_{I}(x)  = 0$ we can write the matter fields in the form, 
\bq
\lb{3.20}
F(x) = F^{+}(x) H\left(\Phi_{I}\right) 
       + F^{-}(x)\left[1 - H\left(\Phi_{I}\right)\right],
\eq
where $F \equiv \left\{\phi, \; \psi, \; B\right\}$,
and $F^{+}\; (F^{-})$ is defined in the region $\Phi_{I} > 0\; (\Phi_{I} < 0)$. Then, we find
that
\bqn
\lb{3.21}
F_{,a}(x) &=& F^{+}_{,a}(x) H\left(\Phi_{I}\right) 
       + F^{-}_{,a}(x)\left[1 - H\left(\Phi_{I}\right)\right],\nb\\
F_{,ab}(x) &=& F^{+}_{,ab}(x) H\left(\Phi_{I}\right) 
       + F^{-}_{,ab}(x)\left[1 - H\left(\Phi_{I}\right)\right]\nb\\
       & & + \left[F_{,a}\right]^{-}\frac{\partial \Phi_{I}(x)}{\partial x^{b}}\; 
\delta\left(\Phi_{I}\right),
\eqn    
where $\left[F_{,a}\right]^{-}$ is defined as that  in Eq.(\ref{3.5}). Projecting
$F_{,a}$ into $n^{a}$ and $e^{a}_{(\mu)}$ directions, we find
\bq
\lb{3.22}
F_{,a} = F_{,\mu} e_{a}^{(\mu)} - F_{,n} n_{a},
\eq
where 
\bq
\lb{3.23}
F_{,n} \equiv n^{a}F_{,a},\;\;
F_{,\mu} \equiv e^{a}_{(\mu)}F_{,a}.
\eq
Then, we have
\bqn
\lb{3.24}
& & \left[F_{,a}\right]^{-} n^{a} = \left[F_{,n}\right]^{-},\nb\\
& & \left[F_{,a}\right]^{-} e^{a}_{(\mu)} = 0.
\eqn
Inserting Eqs.(\ref{3.22})-(\ref{3.24}) into Eq.(\ref{3.21}), we find
\bqn
\lb{3.25}
F_{,ab}(x) &=& F^{+}_{,ab}(x) H\left(\Phi_{I}\right) 
       + F^{-}_{,ab}(x)\left[1 - H\left(\Phi_{I}\right)\right]\nb\\
       & & - \left[F_{,n}\right]^{-}n_{a}n_{b} N_{I} 
       \;  \delta\left(\Phi_{I}\right),
\eqn
where $N_{I} \equiv \sqrt{\left|\Phi_{I,c}\Phi_{I}^{,c}\right|}$, and
\bq
\lb{3.26}
n_{a} = \frac{1}{N_{I}}
\frac{\partial \Phi_{I}(x)}{\partial x^{a}}.
\eq
Substituting Eq.(\ref{3.25}) into Eqs.(\ref{3.ea})-(\ref{3.ed}), we find that the matter
field equations on the branes read,
\bqn
\lb{3.27a}
\left[\phi^{(I)}_{,n}\right]^{-} &=& 
-  \Xi^{(I)} \left(2\kappa_{D}^{2}\epsilon_{I} \frac{\partial{V_{D-1}^{(I)}}}
{\partial{\phi}} + \sigma^{(I)}_{\phi}\right),\\
\lb{3.27b}
\left[\psi^{(I)}_{,n}\right]^{-} &=&
 -  \Xi^{(I)} \left(2\kappa_{D}^{2}\epsilon_{I} \frac{\partial{V_{D-1}^{(I)}}}
{\partial{\psi}} + \sigma^{(I)}_{\psi}\right),\\
\lb{3.27c}
\left[B^{(I)}_{ij,n}\right]^{-} &=& -    \Xi^{(I)} \; \sigma^{(I)}_{ij},\\
\lb{3.27d}
\left[H^{(I)}_{nab}\right]^{-} &=& -   \Xi^{(I)}\; \sigma^{(I)}_{ab},
\eqn
where
\bq
\lb{3.28}
H_{nab} \equiv H_{cab} n^{c},\;\;
\Xi^{(I)} \equiv \frac{1}{N_{I}} \sqrt{\left|\frac{g^{(I)}_{D-1}}{g_{D}}\right|}.
\eq

\begin{figure}
\centering
\includegraphics[width=8cm]{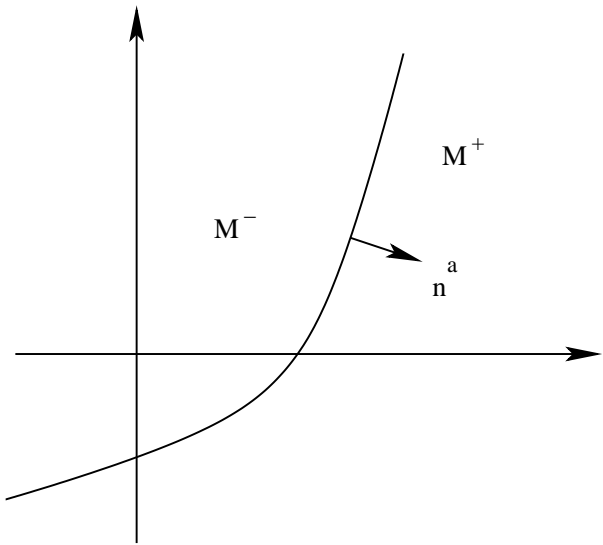}
\caption{The  surface $\Phi_{I}(x) = 0$ divides the spacetimes into two regions, $\Phi_{I}(x) > 0$ 
and $\Phi_{I}(x) < 0$. The normal vector defined by Eq.(\ref{3.26}) points from $M^{-}$ to
$M^{+}$, where $M^{+} \equiv \left\{x: \Phi_{I}(x) > 0\right\}$ and
$M^{-} \equiv \left\{x: \Phi_{I}(x) < 0\right\}$. }
\label{fig1}
\end{figure}
 
This completes our general description for $(D+d)$-dimensional spacetimes of string theory
with  two orbifold branes. Setting $D = d = 5$, we shall obtain the results presented
in \cite{WSVW08}.  from now on we shall restrict ourselves to this case.

\section{Gravitational Coupling in 4-Dimensional Effective Theory and The  Hierarchy Problem}
\renewcommand{\theequation}{4.\arabic{equation}}
\setcounter{equation}{0}
 
 One of the main motivations of the brane worlds is to resolve the long standing 
{\em hierarchy problem}, namely the large difference in magnitudes between the Planck and 
electroweak scales \cite{ADD98,RS1}. In this section, we are going to show explicitly how the 
problem is solved in our current setup.  We first note that in deriving the relation between
the two scales $M_{D}$ and $M_{pl}$, given by Eqs.(\ref{1.2})  and (\ref{1.4}), it was implicitly
assumed that  the 4-dimensional effective Einstein-Hilbert  action $S_{g}^{eff.}$  couples with 
matter directly in the form,
\bq
\lb{4.1}
S_{g}^{eff.} + S_{m} = 
\int{\sqrt{-g}d^{4}\left(-\frac{1}{2\kappa^{2}_{4}} R + {\cal{L}}_{m}\right)},
\eq
from which one obtains the Einstein field equations, $G_{\mu\nu} = \kappa^{2}_{4}\tau_{\mu\nu}$.
In the weak field limit, one arrives at $\kappa^{2}_{4} = 8\pi G/c^{4}$ \cite{d'Inv92}. However, in the 
brane-world scenarios, the coupling between the effective Einstein-Hilbert  action  and matter is 
much more complicated than that given by Eq.(\ref{4.1}). In particular,  the gravitational field equations 
on the branes are given by Eqs.(\ref{3.15})-(\ref{3.17}), which are a second-order polynomial in terms of the 
energy-momentum tensor $\tau_{\mu\nu}$ of the brane. In the weak-field regime, the quadratic
terms are negligible, and the term linear to  $\tau_{\mu\nu}$ dominates. 
Then, under the 
weak-field limit, one can show that $\kappa^{2}_{4}$ defined by Eq.(\ref{3.17}) is related
to the Newtonian constant exactly by  $\kappa^{2}_{4} = 8\pi G/c^{4}$, from which we find
that 
\bq
\lb{4.2}
g_{\kappa} = \frac{6\kappa^{2}_{4}}{\kappa^{4}_{5}}.
\eq
Note that this result is quite general, and applicable to a large class of brane-world scenarios 
\cite{branes}. In the present case, we have $\kappa^{2}_{5} = M^{-3}_{5} = 1/(M^{8}_{10}
R^{5})$, where $R$ is the typical size of the extra dimensions \cite{GWW07}. Then, one find 
that $g_{\kappa}  \simeq 10^{-47}\; GeV^{4}$, that is, to solve the hierarchy problem in the framework
of string theory on $S^{1}/Z_{2}$, the tension of the brane has to be in the 
same order of  the observational cosmological constant $\rho_{\Lambda}^{obs}$.

\section{ Radion Mass }
\renewcommand{\theequation}{5.\arabic{equation}}
\setcounter{equation}{0}

\begin{figure}[tbp]
\includegraphics[width=\columnwidth]{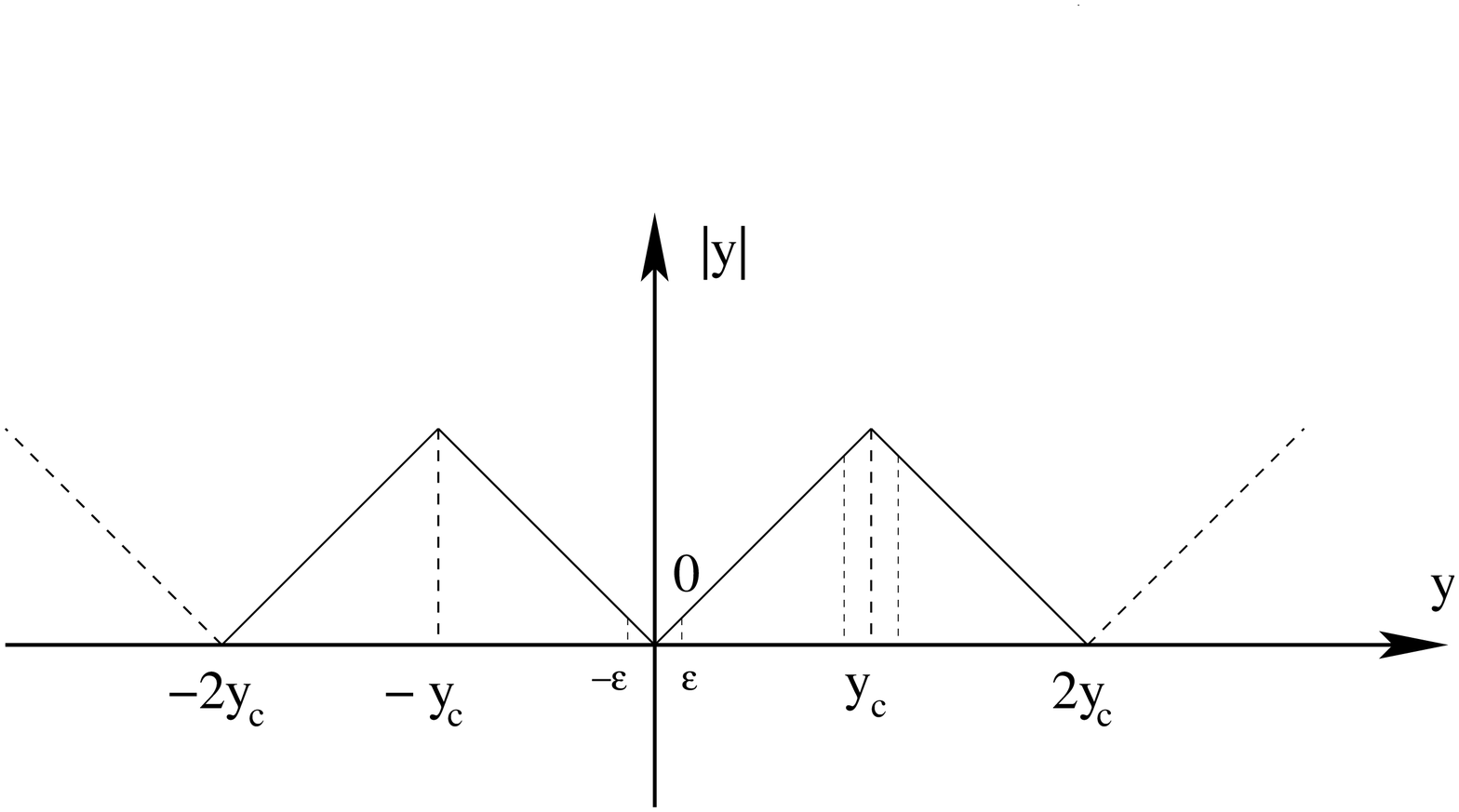}
\caption{The function $\left|y\right|$ appearing in Eq. (\ref{5.1a}). }
\label{fig2}
\end{figure}

In \cite{WS07}, we studied the radion stability using the Goldberger-Wise mechanism
\cite{GW99}, and found that the radion is stable.  To show this claim,   we  considered
the 5-dimensional static metric with a 4-dimensional Poincar\'e symmetry,
\bqn
\lb{5.1a}
ds^{2}_{5} &=& e^{2\sigma(y)}\left(\eta_{\mu\nu}dx^{\mu}dx^{\nu} - dy^{2}\right),\\
\lb{5.1b}
\sigma(y) &=&  \frac{1}{9}\ln\left(\frac{|y| + y_{0}}{L}\right), \nb\\
\phi(y) &=& - \sqrt{\frac{25}{54}}\; \ln\left(\frac{|y| + y_{0}}{L}\right) + \phi_{0},\nb\\
\psi(y) &=& -  \sqrt{\frac{5}{18}}\; \ln\left(\frac{|y| + y_{0}}{L}\right)
+ \psi_{0}, \nb\\
B_{ij} &=& 0 = B_{ab},
\eqn
where $|y|$ is defined as in Fig. \ref{fig2},  ${L}$ and $y_{0}$ are positive constants, and
\bq
\lb{5.1c}
\psi_{0} \equiv \sqrt{\frac{2}{5}}\left(\ln\left(\frac{2}{9{L}^{2}V^{0}_{(5)}}\right)
- \frac{5}{\sqrt{6}}\phi_{0}\right).
\eq
Then, it can be shown that  the above solution satisfies the  gravitational and matter 
field equations both outside and on the branes, for any given potentials of the branes
for $\tau^{(I)}_{\mu\nu} = 0$. For the detail, we refer readers to \cite{WS07}.

To study the radion stability and  mass,   it is found convenient to introduce 
the proper distance $Y$, defined by \cite{WS07}
\bq
\lb{5.2a}
Y = \left(\frac{9L}{10}\right)\left\{\left(\frac{y+y_{0}}{L}\right)^{10/9} 
- \left(\frac{y_{0}}{L}\right)^{10/9}\right\}.
\eq
Then, in terms of $Y$, the static solution (\ref{5.1a}) can be written as
\bq
\lb{5.1aa}
ds^{2}_{5} = e^{-2A(Y)}\eta_{\mu\nu}dx^{\mu}dx^{\nu} - dY^{2},
\eq
with 
\bqn
\lb{5.1ab}
A(Y) &=&  -\frac{1}{10}\ln\left\{\left(\frac{10}{9L}\right) \left(|Y| 
+ Y_{0}\right)\right\},\nb\\
\phi(Y) &=&  -\sqrt{\frac{3}{8}}\ln\left\{\left(\frac{10}{9L}\right) 
\left(|Y| + Y_{0}\right)\right\}
+ \phi_{0},\nb\\
\psi(Y) &=&  -\frac{3}{\sqrt{40}}\ln\left\{\left(\frac{10}{9L}\right) 
\left(|Y| + Y_{0}\right)\right\}\nb\\
& &
+ \psi_{0},
\eqn
where $|Y|$ is defined  also as that of Fig. \ref{fig2},
with  
\bqn
\lb{5.2b}
Y_{0} &\equiv& \left(\frac{9L}{10}\right)\left(\frac{y_{0}}{L}\right)^{10/9}, \nb\\
Y_{c} &\equiv& \left(\frac{9L}{10}\right)\left\{\left(\frac{y_{c} + y_{0}}{L}\right)^{10/9} 
- \left(\frac{y_{0}}{L}\right)^{10/9}\right\},
\eqn
and $Y_{2} = 0, \; Y_{1} = Y_{c}$.

\begin{figure}
\centering
\includegraphics[width=8cm]{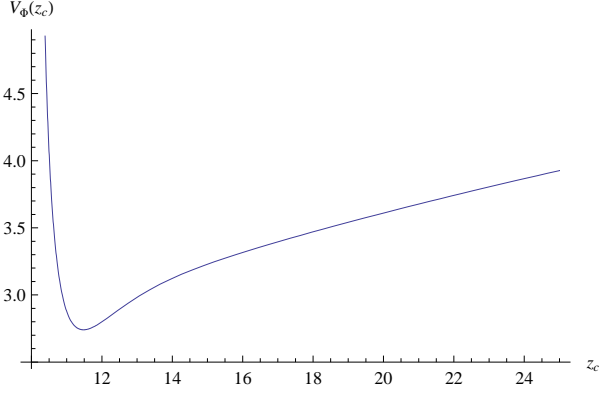}
\caption{The  potential defined by Eq.(\ref{5.4}) in the  limit of large $v_{I}$ and $y_{0}$.
In this particular plot, we choose $(z_{0}, \;  v_{1}, \;  v_{2}) = (10, \;  1.0,\; 0.1)$. }
\label{fig3}
\end{figure}

Following \cite{GW99}, in cite{WS07} we   considered a massive scalar field $\Phi$ 
in the background of the spacetime described above and found that the radion potential
is given by, 
\bqn
\lb{5.5g}
V_{\Phi}\left(Y_{c}\right) &\equiv&  - \int_{0+\epsilon}^{Y_{c}- \epsilon}{dY \sqrt{\left|g_{5}\right|} 
\left(\left(\nabla\Phi\right)^{2} - m^{2}\Phi^{2}\right)}\nb\\
& & + \sum_{I=1}^{2}{ \alpha_{I}  \int_{Y_{I} -\epsilon}^{Y_{I} + \epsilon}
{dY \sqrt{\left|g_{4}^{(I)}\right|} 
\left(\Phi^{2} - v^{2}_{I}\right)^{2}}}\nb\\
& & \;\;\;\;\;\;\; \;\;\;\;\;\;\; \times \delta\left(Y-Y_{I}\right) \nb\\
&=& \left. e^{-4A(Y)}\Phi(Y)\Phi'(Y)\right|^{Y_{c}}_{0}\nb\\
& &
+ \sum_{I=1}^{2}{\alpha_{I} \left(\Phi^{2}_{I} - v^{2}_{I}\right)^{2}e^{-4A(Y_{I})}}.
\eqn	 
In the limits that $\alpha_{I}$'s are very large  and  
$ m Y_{0} \gg 1$ \cite{GW99}, we found  
\bqn
\lb{5.4}
V_{\Phi}\left(Y_{c}\right) &=& \left(\frac{10Y_{0}}{9L}\right)^{2/5}
\frac{ M}{\sinh\left(z_{c} - z_{0}\right)}
   \left\{- 2v_{1}v_{2} \right.\nb\\
   & & \left. +
   \left(v_{1}^{2} + v_{2}^{2}\right)\cosh\left(z_{c} - z_{0}\right) \right\},
\eqn
from which we find that 
\bqn
\lb{5.6}
\frac{\partial V_{\Phi}\left(Y_{c}\right)}{\partial Y_{c}} &=&
\left(\frac{10Y_{0}}{9L}\right)^{2/5}
\frac{2v_{1}v_{2}M}{\sinh^{2}\left(z_{c} - z_{0}\right)}
   \left\{ \cosh\left(z_{c} - z_{0}\right) \right.\nb\\
   & & \left. - \frac{v_{1}^{2} + v_{2}^{2}}{2v_{1}v_{2}}\right\},
\eqn
where $z_{c} - z_{0} = MY_{c}$.  Figs. \ref{fig3}  shows the potential for 
$(z_{0}, \;  v_{1}, \;  v_{2}) 
= (10, \;  1.0,\;0.1)$.  Clearly, $V_{\Phi}\left(Y_{c}\right)$ has a minimum at
\bq
\lb{5.7}
Y^{min}_{c}  = \frac{1}{M} \cosh^{-1}\left(\frac{v_{1}^{2} + v_{2}^{2}}{2v_{1}v_{2}}\right),
\eq
for which we have
\bq
\lb{5.6a}
\left. \frac{\partial^{2} V_{\Phi}\left(Y_{c}\right)}{\partial Y_{c}^{2}}\right|_{Y_{c}
= Y^{min}_{c}} = \left(\frac{10Y_{0}}{9L}\right)^{2/5}
\frac{4v_{1}v_{2}M^{3}}{\left|v_{1}^{2} - v_{2}^{2}\right|}.
\eq

As shown in \cite{GW99,WGW08}, the radion field $\varphi$ is related to the proper distance
$Y_{c}$ between the two branes by
\bq
\lb{5.7a}
\varphi\left(Y_{c}\right) = \sqrt{12f\left(Y_{c}\right)},
\eq
where
\bqn
\lb{5.8}
f &\equiv& \frac{1}{\kappa^{2}_{5}} 
\int_{0}^{Y_{c}}{e^{-2A(Y)}dY} = \frac{5L}{6\kappa^{2}_{5}}  \left(\frac{10}{9}\right)^{1/5}\nb\\
&&  \times
\left\{\left(\frac{Y_{c} + Y_{0}}{L}\right)^{6/5}  -  \left(\frac{Y_{0}}{L}\right)^{6/5}\right\}.    
\eqn
Then, we find that
\bqn
\lb{5.9}
m_{\varphi}^{2} &=& \frac{1}{2}\left. \frac{\partial^{2} V_{\Phi}\left(Y_{c}\right)}
{\partial \varphi^{2}}\right|_{Y_{c}
= Y^{min}_{c}} = \left(\frac{10Y_{0}}{9L}\right)^{1/5}\frac{2M^{5}}{3M^{3}_{5}}\nb\\
& & \times \tilde{v}^{2}_{1}\tilde{v}^{2}_{2}\left|\frac{\ln(\tilde{v}_{1}/\tilde{v}_{2})}
{\tilde{v}_{1}^{2} - \tilde{v}_{2}^{2}}\right|,
\eqn
where $v_{i} = M^{3/2}\tilde{v}_{i}$. Since $v_{i}$ has the dimension $[m]^{3/2}$, we can see
that $\tilde{v}_{i}$ is dimensionless. in addition, $M$ and $v_{i}$ are all
 5-dimensional quantities,
we expect that $M \sim M_{5}$ and $\tilde{v}_{i} \sim {\cal{O}}(1)$. Without introducing new
hierarchy, we also expect that $\left(Y_{0}/L\right)^{1/5} \sim {\cal{O}}(1)$ and
$Y_{c}/Y_{0} \sim {\cal{O}}(1)$. Then, from 
Eq.(\ref{5.9}) we find 
\bq
\lb{5.10}
m_{\varphi}  \simeq M_{5}  = \left(\frac{M_{10}}{M_{pl}}\right)^{8/3}
\left(\frac{R}{l_{pl}}\right)^{5/3}M_{pl}. 
\eq
For $M_{10} \sim \; TeV$ and $R \sim \; 10^{-22}\; m$, we find that $m_{\varphi}  \simeq
10^{-2}\; GeV$, which is much large than the experimental limit $m_{\varphi} > 10^{-3} \;
eV$ \cite{RS1}.

\section{ Localization of Gravity and 4D Effective Newtonian Potential}
\renewcommand{\theequation}{6.\arabic{equation}}
\setcounter{equation}{0}

To study the localization of gravity and the four-dimensional effective gravitational 
potential, in this section let us consider small fluctuations $h_{ab}$ of the 5-dimensional 
static metric with a 4-dimensional Poincar\'e symmetry, given by Eqs.(\ref{5.1a}) and (\ref{5.1b})
in its conformally flat form. 

\subsection{Tensor Perturbations and the KK Towers}

\begin{figure}
\centering
\includegraphics[width=8cm]{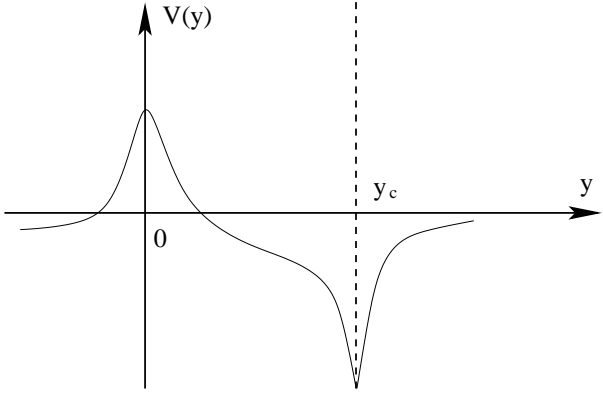}
\caption{The  potential defined by Eq.(\ref{7.7}). }
\label{fig5}
\end{figure}

Since such tensor perturbations are not coupled with scalar ones
\cite{GT00}, without loss of generality, we can set the perturbations of the scalar fields
$\phi$ and $\psi$ to zero, i.e., $\delta \phi = 0 = \delta \psi$. We shall choose the gauge
\bq
\lb{7.1}
h_{ay} = 0, \;\;\; 
h^{\lambda}_{\lambda} = 0 = \partial^{\lambda}h_{\mu\lambda}.
\eq
Then, it can be shown that \cite{Csaki00}
\bqn
\lb{7.2}
\delta{G}^{(5)}_{ab} &=& - \frac{1}{2}\Box_{5}h_{ab}
- \frac{3}{2} \left\{\left(\partial_{c}\sigma\right)\left(\partial^{c}h_{ab}\right)\right.\nb\\
& & \left.
- 2\left[\Box_{5}\sigma + \left(\partial_{c}\sigma\right)
\left(\partial^{c}\sigma\right)\right]h_{ab}\right\}, \nb\\
\kappa^{2}_{5}\delta{T}^{(5)}_{ab} &=& \frac{1}{4}\left({\phi'}^{2} + {\psi'}^{2}
+ 2 e^{2\sigma}V_{5}\right)h_{ab},\nb\\
\delta{T}^{(4)}_{\mu\nu} &=&  \left(\tau^{(I)}_{(\phi, \psi)} + 2\rho^{(I)}_{\Lambda}\right) 
e^{2\sigma(y_{I})}h_{\mu\nu}(x, y_{I}),
\eqn
where $\Box_{5} \equiv \eta^{ab}\partial_{a}\partial_{b}$ and 
$\left(\partial_{c}\sigma\right)\left(\partial^{c}h_{ab}\right)
\equiv \eta^{cd} \left(\partial_{c}\sigma\right)\left(\partial_{d}h_{ab}\right)$,
with $\eta^{ab}$ being the five-dimensional Minkowski metric. Substituting the above
expressions into the Einstein field equations (\ref{3.3}) with $D = 5$, 
and noticing that
\bq
\lb{7.2a}
\left|\frac{g^{(I)}_{4}}{g_{5}}\right|^{1/2} = e ^{-\sigma(y_{I})},
\eq
we find that in the present case there is only one independent equation, given by
\bq
\lb{7.3}
\Box_{5}h_{\mu\nu} + 3 \left(\partial_{c}\sigma\right)\left(\partial^{c}h_{\mu\nu}\right) = 0,
\eq
which can be further cast in the form,
\bq
\lb{7.4}
\Box_{5}\tilde{h}_{\mu\nu} + \frac{3}{2} \left(\sigma''
+ \frac{3}{2}  \sigma'\right)  \tilde{h}_{\mu\nu} = 0,
\eq
where $h_{\mu\nu} \equiv e^{-3\sigma/2}\tilde{h}_{\mu\nu}$. Setting
\bqn
\lb{7.5}
& & \tilde{h}_{\mu\nu}(x, y) = \hat{h}_{\mu\nu}(x) \psi(y),\nb\\
& & \Box_{5} =  \left(\Box_{4} - \nabla^{2}_{y}\right) = 
 \left(\eta^{\mu\nu}\partial_{\mu}\partial_{\nu} - \partial^{2}_{y}\right),\nb\\
& & \Box_{4}\hat{h}_{\mu\nu}(x) = - m^{2}\hat{h}_{\mu\nu}(x),
\eqn
we find that Eq.(\ref{7.3}) takes the form of the schr\"odinger equation,
\bq
\lb{7.6}
\left(- \nabla^{2}_{y} + V\right)\psi = m^{2}\psi,
\eq
where
\bqn
\lb{7.7}
V &\equiv& \frac{3}{2}\left(\sigma'' + \frac{3}{2}{\sigma'}^{2}\right)\nb\\
&=& -  \frac{5}{36\left(\left|y\right| + y_{0}\right)^{2}} 
      +  \frac{\delta\left(y\right)}{3y_{0}}\nb\\
& & - \frac{\delta\left(y-y_{c}\right)}{3\left(y_{c} + y_{0}\right)}.
\eqn
From the above expression we can see clearly that the potential has a delta-function well
at $y = y_{c}$, which is responsible for the localization of the graviton on this brane.
In contrast, the potential has a delta-function barrier at $y = 0$, which makes the gravity
delocalized on the  $y = 0$ brane.  Fig. \ref{fig5} shows the potential schematically.

Introducing the operators,
\bq
\lb{7.8}
Q \equiv \nabla_{y} - \frac{3}{2}\sigma',\;\;\;
Q^{\dagger} \equiv - \nabla_{y} - \frac{3}{2}\sigma',
\eq
Eq.(\ref{7.6}) can be written in the form of a supersymmetric quantum mechanics problem,
\bq
\lb{7.9} 
Q^{\dagger}\cdot Q\psi = m^{2}\psi.
\eq
It should be noted that Eq.(\ref{7.9}) itself does not quarantee that the operator
$Q^{\dagger}\cdot Q$ is Hermitian, because now it is defined only on a finite interval,
$y \in [0, y_{c}]$. To ensure its Hermiticity, in addition to writing the differential
equation  in the Shr\"odinger form, one also needs to show that it has Hermitian boundary
conditions, which can be formulated as \cite{Csaki01}
\bqn
\lb{7.9a}
{\psi'}_{n}(0) \psi_{m}(0)  &-& \psi_{n}(0) {\psi'}_{m}(0) =
{\psi'}_{n}\left(y_{c}\right) \psi_{m}\left(y_{c}\right)  
\nb\\
&-& \psi_{n}\left(y_{c}\right) {\psi'}_{m}\left(y_{c}\right),
\eqn
for any two solutions of Eq.(\ref{7.9}). To show that in the present case this condition
is indeed satisfied, let us consider  the boundary conditions at $y = 0$ and 
$y = y_{c}$. Integration of Eq.(\ref{7.6}) in the neighbourhood of $y = 0$ and $y= y_{c}$
yields, respectively, the conditions,     
\bqn
\lb{7.12a}
\lim_{y \rightarrow y^{-}_{c}}{\psi'(y)} &=&
\frac{1}{6\left(y_{c} + y_{0}\right)} \lim_{y \rightarrow y^{-}_{c}}{\psi(y)},\\ 
\lb{7.12b}
\lim_{y \rightarrow 0^{+}}{\psi'(y)} &=&
\frac{1}{6 y_{0}} \lim_{y \rightarrow 0^{+}}{\psi(y)}.
\eqn
Note that in writing the above equations we had used the $Z_{2}$ symmetry of the 
wave function $\psi$. Clearly, any solution of Eq.(\ref{7.6}) that satisfies the above boundary
conditions also satisfies Eq.(\ref{7.9a}). That is, the operator $Q^{\dagger}\cdot 
Q$ defined by Eq.(\ref{7.8}) is indeed a positive definite Hermitian  operator. Then, by
the usual theorems we can see that  all eigenvalues $m_{n}^{2}$ are non-negative, and their corresponding 
wave functions $\psi_{n}(y)$ are orthogonal to each other and form a complete basis. Therefore, 
{the background is gravitationally stable  in our current setup}. 

\subsubsection{Zero Mode}

The four-dimensional gravity is given by the existence of the normalizable
zero mode, for which the corresponding wavefunction is given by
\bq
\lb{7.10}
\psi_{0} (y) = N_{0}\left(\frac{|y| + y_{0}}{L}\right)^{1/6}, 
\eq
where $N_{0}$ is the normalization factor, defined as
\bq
\lb{7.11}
N_{0} \equiv 2\left\{3L\left[\left(\frac{y_{c} + y_{0}}{L}\right)^{4/3}
- \left(\frac{y_{0}}{L}\right)^{4/3}\right]\right\}^{-1/2}.
\eq	
Eq.(\ref{7.10}) shows clearly that the wavefunction is increasing as $y$ increases
from $0$ to $y_{c}$. Therefore, the gravity is indeed localized near the $y = y_{c}$
brane. 

\subsubsection{Non-Zero Modes}

\begin{figure}
\centering
\includegraphics[width=8cm]{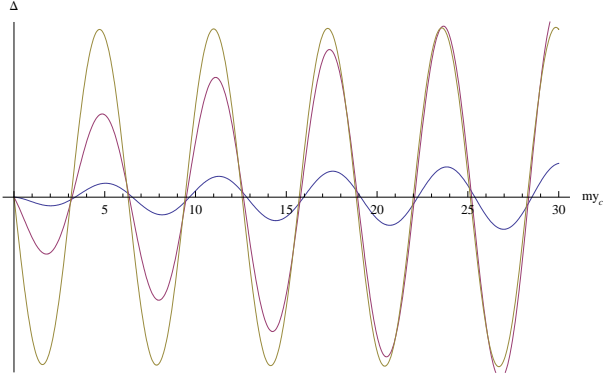}
\caption{The function of $ \Delta$  defined by Eq.(\ref{7.18}) for
$z_{0} = my_{0}
= 0.01, \; 1.0, \; 1000$, respectively.   }
\label{roots}
\end{figure}

In order to have localized four-dimensional gravity, we require that the corrections
to the Newtonian law from the non-zero modes, the KK modes, of Eq.(\ref{7.6}), be
very small, so that they will not lead to contradiction with observations. 
To solve Eq.(\ref{7.6}) outside of the two branes, it is found convenient to introduce
the quantities,
\bq
\lb{7.13}
\psi(y) \equiv z^{1/2} \; u(z),\;\;\;
z \equiv m\left(y + y_{0}\right).
\eq
Then, in terms of $z$ and $u(z)$, Eq.(\ref{7.6}) takes the form,
\bq
\lb{7.14}
z^{2}\frac{d^{2}u}{dz^{2}} + z\frac{du}{dz}
+ \left(z^{2} - \nu^{2}\right) u = 0,
\eq
but now with $\nu = 1/3$. Eq.(\ref{7.14}) is the standard Bessel equation \cite{AS72},
which have two independent  solutions $J_{\nu}(z)$ and $Y_{\nu}(z)$. Therefore, the
general solution of Eq.(\ref{7.6}) are given by
\bq
\lb{7.15}
\psi = z^{1/2}\left\{c J_{\nu}(z) + d Y_{\nu}(z)\right\},
\eq
where $c$ and $d$ are the integration constants, which will be determined from the 
boundary conditions given by Eqs.(\ref{7.12a}) and (\ref{7.12b}). Setting
\bqn
\lb{7.16}
\Delta_{11} &\equiv& 2J_{\nu}\left(z_{c}\right) - 3z_{c}J_{\nu+1}\left(z_{c}\right),\nb\\
\Delta_{12} &\equiv& 2Y_{\nu}\left(z_{c}\right) - 3z_{c}Y_{\nu+1}\left(z_{c}\right),\nb\\
\Delta_{21} &\equiv& 2J_{\nu}\left(z_{0}\right) - 3z_{0}J_{\nu+1}\left(z_{0}\right),\nb\\
\Delta_{22} &\equiv& 2Y_{\nu}\left(z_{0}\right) - 3z_{0}Y_{\nu+1}\left(z_{0}\right), 
\eqn
we find that Eqs.(\ref{7.12a}) and (\ref{7.12b}) can be cast in the form,
\bq
\lb{7.17}
\left(\matrix{\Delta_{11} & \Delta_{12}\cr
\Delta_{21} & \Delta_{22}\cr}\right)\left(\matrix{c  \cr
d\cr}\right) = 0.
\eq
It has no trivial solutions only when 
\bq
\lb{7.18}
\Delta \equiv {\mbox{det}}\left(\Delta_{ij}\right) = 0.
\eq
Fig. \ref{roots} shows the solutions of $\Delta = 0$ for $z_{0} = my_{0}
= 0.01, \; 1.0, \; 1000$, respectively, where the root $my_{c} = 0$ is
the zero-mode, discussed in the last sub-section. Thus, in the rest of this 
sub-section, we shall not consider it. From this figure, two remarkable 
features emerge: (1) The spectrum of the KK towers is discrete. (2) The KK
modes weakly depend on the specific values of $z_{0}$. 

Table I shows the first three modes $m_{n}\; (n = 1, 2, 3)$ for 
$z_{0} = 0.01, \; 1.0, \; 1000$, 
from which we can see that to find $m_{n}$ it is sufficient to consider only 
the case where $z_{0} \gg 1$. 

\begin{table} 
\begin{tabular}{|c|c|c|c|} \hline
\lb{table1}
$z_{0}$ & $m_{1}y_{c}$ &  $m_{2}y_{c}$ &  $m_{3}y_{c}$ \\ \hline
0.01 & 3.37 & 6.52 & 9.67\\ \hline
1.0 & 3.20 & 6.35 & 9.50\\ \hline
1000 & 3.14 & 6.28 & 9.42\\ \hline
\end{tabular}
\caption{The first three modes $m_{n}\; (n = 1, 2, 3)$ for $z_{0}  
= 0.01, \; 1.0, \; 1000$, respectively.}
\end{table}
 
When $z_{0} \gg 1$ we find that $z_{c} = z_{0} + my_{c} \gg 1$, and that
\cite{AS72}
\bqn
\lb{7.19}
J_{\nu}(z) &\simeq& - Y_{\nu + 1}(z) \simeq \sqrt{\frac{2}{\pi z}}
\cos\left(z - \frac{5}{12}\pi\right),\nb\\
Y_{\nu}(z) &\simeq&  J_{\nu + 1}(z) \simeq \sqrt{\frac{2}{\pi z}}
\sin\left(z - \frac{5}{12}\pi\right).
\eqn
Inserting the above expressions into Eqs.(\ref{7.16}) and (\ref{7.18}), we obtain
\bqn
\lb{7.20}
\Delta &=& - \sqrt{\frac{4}{\pi^{2} z_{0}z_{c}}}
\left\{6\left(z_{c} - z_{0}\right)\cos\left(z_{c} - z_{0}\right)\right.\nb\\
& & \left.
+ \left(4 + 9 z_{0}z_{c}\right)\sin\left(z_{c} - z_{0}\right)\right\},
\eqn
whose roots are given by
\bq
\lb{7.21}
\tan\left(z_{c} - z_{0}\right) = - \frac{6\left(z_{c} - z_{0}\right)}{4 + 9 z_{0}z_{c}}.
\eq
From this equation, we can see that $m_{n}$  satisfies the bounds
\bq
\lb{7.22}
\left(n - \frac{1}{2}\right) \frac{\pi}{y_{c}} < m_{n}  <    
\frac{n\pi}{y_{c}},\; (n = 1, 2, 3, ...).
\eq
Combining the above expression with Table I, we find that $m_{n}$
is well approximated by
\bq
\lb{7.23}
m_{n} \simeq n\pi \left(\frac{l_{pl}}{y_{c}}\right)M_{pl}, 
\eq
For $z_{0} \gg 1$. In particular, we have
\bqn
\lb{7.24}
m_{1} &\simeq& 3.14 \times \left(\frac{10^{-19}\; {\mbox{m}}}{y_{c}}\right)\; {\mbox{TeV}}\nb\\
&\simeq&  3.14\times \cases{1\; {\mbox{TeV}}, & $y_{c} \simeq 10^{-19} \; {\mbox{m}}$,\cr
10^{-2} \; {\mbox{eV}}, & $y_{c} \simeq 10^{-5} \; {\mbox{m}}$,\cr
10^{-4} \; {\mbox{eV}}, & $y_{c} \simeq 10^{-3} \; {\mbox{m}}$.\cr}
\eqn

It should be noted that the   mass $m_{n}$ calculated above is measured
by the observer with the metric $\eta_{\mu\nu}$. However, since the warped factor
$e^{\sigma(y)}$ is different from one at $y = y_{c}$, the physical mass on the visible brane 
should be given by \cite{RS1}
\bq
\lb{phycialmass1}
m^{obs}_{n} = e^{-\sigma\left(y_{c}\right)}m_{n}
= \left(\frac{y_{c} + y_{0}}{L}\right)^{-1/9} m_{n}.
\eq
Without introducing any new hierarchy, we expect that 
$\left[({y_{c} + y_{0})}/{L}\right]^{-1/9} \simeq {\cal{O}}(1)$. As a result, 
we have 
\bq
\lb{phycialmass2}
m^{obs}_{n}  
= \left(\frac{y_{c} + y_{0}}{L}\right)^{-1/9} m_{n} \simeq m_{n}.
\eq

For each $m_{n}$ that satisfies Eq.(\ref{7.18}), the wavefunction $\psi_{n}(z)$ is given by
\bqn
\lb{7.25}
\psi_{n}(z) &=& N_{n}z^{1/2}\left\{\Delta_{12}\left(m_{n}, y_{c}\right) J_{\nu}(z) \right.\nb\\
& & \left.
- \Delta_{11}\left(m_{n}, y_{c}\right) Y_{\nu}(z)\right\},
\eqn
where $N_{n} \equiv N_{n}\left(m_{n}, y_{c}\right)$ is the normalization factor, so that
\bq
\lb{7.26}
\int^{y_{c}}_{0}{\left|\psi_{n}(z)\right|^{2} dy} = 1.
\eq

\subsection{4D Newtonian Potential and Yukawa Corrections}

To calculate the four-dimensional effective Newtonian potential and its corrections,
let us consider two point-like sources of masses $M_{1}$ and $M_{2}$, located on the 
brane at $y = y_{c}$. Then, the discrete eigenfunction $\psi_{n}(z)$ of mass $m_{n}$
has an Yukawa correction to the four-dimensional gravitational potential between 
the two particles \cite{BS99,Csaki00}
\bq
\lb{7.27}
U(r) = G_{4}\frac{M_{1}M_{2}}{r} 
       + \frac{M_{1}M_{2}}{M^{3}_{5}r} 
       \sum^{\infty}_{n =1}{e^{-m_{n}r} \left|\psi_{n}(z_{c})\right|^{2}},
\eq
where $\psi_{n}(z_{c})$ is given by Eq.(\ref{7.25}). When $z_{0} = m_{n}y_{0}
\gg 1$, from Eqs.(\ref{7.19}),  (\ref{7.25}) and (\ref{7.26}) we find that
\bqn
\lb{7.28}
N_{n}  &\simeq& \sqrt{\frac{\pi^{2}}{18z_{c}y_{c}}},\nb\\
\psi_{n}(z_{c}) &\simeq& \sqrt{\frac{2}{y_{c}}}.
\eqn
Then, we obtain  
\bq
\lb{7.29}
\delta_{1}(r) \simeq  \left(\frac{10^{28} \; {\mbox{m}}}{y_{c}}\right)  
 e^{-\frac{\pi r}{2y_{c}}}.
\eq
Clearly, for $y_{c} \simeq 10^{-19} \; {\mbox{m}}$ and
$r \simeq 10 \; \mu{\mbox{m}}$, we have $ \delta_{1}(r) \ll 1$, and the corresponding
Yukawa corrections are negligible.

\section{Conclusions} 

\renewcommand{\theequation}{7.\arabic{equation}}
\setcounter{equation}{0}

In this paper, we have  systematically studied the brane worlds of string theory
on $S^{1}/Z_{2}$. Starting with the toroidal compactification of 
the Neveu-Schwarz/Neveu-Schwarz sector  in (D+d) dimensions, in Sec. II.A
we have first obtained an effective $D$-dimensional action given by Eq.(\ref{2.16})
for  non-vanishing dilaton field and flux with an effective potential given by 
Eq.(\ref{2.17}). Then, in Sec. II.B we have compactified one of the $(D-1)$ spatial
dimensions by adding two orbifold branes as the boundaries of the spacetime
along the compactified dimension. 

Variations of the total action with the metric and matter fields yield, respectively,
the gravitational and matter field equations. This has been done in Sec. III
and given by Eqs.(\ref{3.3})-(\ref{3.ee}).
Dividing the whole set of the  field equations into two groups, one holds outside 
the two branes, and the other holds on  them, in Sec. III.A we have first written down
the field equations outside the two branes, Eqs.(\ref{3.ef})-(\ref{3.ej}), while in 
Sec. III.B, we have written down explicitly the general gravitational field 
equations on each of the two branes,  Eqs. (\ref{3.15})-(\ref{3.17}), by 
combining  the Gauss-Codacci and Lanczos equations. On the other hand, by using 
the distribution theory,  we  have also been able to write down 
the matter field equations on the branes in terms of the discontinuities
of the first derivatives of the matter fields, Eqs. (\ref{3.27a})-(\ref{3.28}).

In the study of orbifold branes, one of the most attractive features is that it may 
resolve the long standing hierarchy problem. In Sec. IV, we have shown explicitly
how it can be solved in the current setup. The mechanism is essentially the combination 
of the ADD large extra dimension \cite{ADD98} and RS warped factor
\cite{RS1} mechanisms together with the tension coupling scenario \cite{Cline99}. 
In order to solve the hierarchy problem in the current setup,  the tensions of the 
branes are required to be in the order of the cosmological constant. 

Another important issue in brane worlds is the radion stability and radion mass 
\cite{branes}. Previously, we showed that the radion is stable \cite{WS07}.
In this paper, we have devoted Sec. V to study the radion mass. With some very 
conservative arguments, we have found that the radion mass is of the order of 
$10^{-2}\; GeV$, which is by far beyond its current observational constraint, 
$m_{\varphi} > 10^{-3}\; eV$. 
 
In Sec. VI we have also shown that the gravity is localized on the visible (TeV) brane, 
in contrast to the RS1 model in which the gravity is localized on the Planck 
(hidden) brane \cite{RS1}. In addition, the spectrum of the gravitational KK
modes is discrete, and given explicitly by Eq.(\ref{7.23}), which can be of the order
of TeV. The corrections to the 4D Newtonian potential from the higher order 
gravitational KK modes are exponentially  suppressed and can be safely neglected
[cf. Eq.(\ref{7.27})].

In Appendix, we have also presented a systematical and pedagogical study of the Gauss-Codacci 
equations and Israel's junction conditions across a surface, which can be either spacelike 
or timelike,  in higher dimensional spacetimes.

It should be noted that, when studied the radion stability, we have ignored the backreaction
of the perturbations. Although it is expected that the main results obtained here will be
continuously valid even after taking such backreaction into acocunt, as what exactly happened
in the Randall-Sundrum model \cite{radion}, it would be very interesting to show explicitly
that this is indeed the case.  

Other important issues that have not been addressed in this paper include  the constraints 
from the solar system tests \cite{solar}, and linear perturbations  in the current setup.

\begin{acknowledgments}

 The authors thank Qiang Wu for the help of preparing some of the figures.   AW would also like to 
 thank T. Ali, G. Cleaver, M. Devin, Y. Huang,  K. Koyama, D. Matravers, A. Papazoglou, Y.-S. Song, 
 and D. Wands for valuable discussions. He also would also like to express his gratitude to the Institute 
 of Cosmology and Gravitation (ICG) for hospitality. NOS was  partially supported by CNPq, while AW 
 was partially supported by  NSFC under Grant,   No. 10703005 $\&$ No. 10775119.
  
\end{acknowledgments}

\section*{Appendix: Gauss-Codacci Equations and Israel's Junction Conditions in Higher 
Dimensional Spacetimes}

In this appendix, we shall present a systematic and pedagogical
study of the Gauss-Codacci equations and Israel's junction conditions across a surface,
where the metric coefficients are only continuous, i.e., $C^{0}$) in higher dimensional  
spacetimes.

\subsection{Notations and Conventions}
\renewcommand{\theequation}{A.\arabic{equation}}
\setcounter{equation}{0}

We shall closely follow  notations and conventions of d'Inverno \cite{d'Inv92}. The 
metric is given by
\bq
\lb{A.1}
ds^{2}_{D} = g_{ab}\left(x^{c}\right) dx^{a} dx^{b},
\eq
with the signature \cite{footnote1},
\bq
\lb{A.1a}
{\rm sign}\left(g_{ab}\right)  = \left\{+, -, -, ..., -\right\}.
\eq
We shall use the lowercase Latin indices, such as, 
$a,\; b,\; c$, to run from $0$ to $D-1$, and the Greek indices, such as, $\mu,\; 
\nu,\; \lambda$, to run from $0$ to $D-2$. The Riemann tensor is defined 
by \cite{footnote2},
\bq
\lb{A.2}
\left(\nabla_{c}\nabla_{d} - \nabla_{d}\nabla_{c}\right)X^{a} = 
 {}^{(D)}R^{a}_{bcd}X^{b},
\eq
where $\nabla_{a}$ denotes the covariant derivative with respect to 
$g_{ab}$. In terms of the Christoffel symbols, it is given by  
\bqn
\lb{A.3}
{}^{(D)}R^{a}_{bcd} &\equiv& {}^{(D)}\Gamma^{a}_{bd,c} - 
{}^{(D)}\Gamma^{a}_{bc,d} 
+ {}^{(D)}\Gamma^{a}_{ce}\; {}^{(D)}\Gamma^{e}_{bd} \nb\\
& & - 
{}^{(D)}\Gamma^{a}_{de}\; {}^{(D)}\Gamma^{e}_{bc},
\eqn
where
\bq
\lb{A.4} 
{}^{(D)}\Gamma^{a}_{bc} = \frac{1}{2}g^{ad}\left(g_{dc,b} + 
g_{bd,c} - g_{bc,d}\right),
\eq
and $g_{ab,c} \equiv \partial g_{ab}/\partial x^{c}$, etc. 
The Ricci and  Einstein tensors are defined as  
\bqn
\lb{A.5}
R^{(D)}_{ab} &\equiv& {}^{(D)}R^{c}_{acb}  
= {}^{(D)}\Gamma^{c}_{ab,c} - {}^{(D)}\Gamma^{c}_{ac,b} \nb\\
& & + {}^{(D)}\Gamma^{c}_{ce}\; {}^{(D)} \Gamma^{e}_{ab}  
- {}^{(D)} \Gamma^{c}_{be}\; {}^{(D)} \Gamma^{e}_{ac},\nb\\
 G^{(D)}_{ab} &\equiv& R^{(D)}_{ab} - \frac{1}{2}g_{ab} R^{(D)},
\eqn
where
\bq
\lb{A.6}
R^{(D)} \equiv R^{(D)}_{ab} g^{ab}.
\eq
The Weyl tensor is defined as
\bqn
\lb{A.8}
 C^{(D)}_{abcd} &=& R^{(D)}_{abcd} 
 + \frac{1}{D-2}\left(g_{ad}\; 
 R^{(D)}_{bc}   \right.\nb\\
& &  + g_{bc}\; R^{(D)}_{ad}
- g_{ac}\;  R^{(D)}_{bd} \nb\\
& & \left.
- g_{bd}\;  R^{(D)}_{ac}\right)\nb\\
& & + \frac{1}{(D-1)(D-2)}\left(g_{ac}g_{bd}\right.\nb\\
& & \left. -g_{ad}g_{bc}\right) \;R^{(D)}.
\eqn

In this paper, we also use the convention,
\bq
\lb{A.8a}
{}^{(D)}X \equiv X^{(D)}.
\eq

\subsection{Gauss and Codacci Equations}

\renewcommand{\theequation}{B.\arabic{equation}}
\setcounter{equation}{0}

Assume that $M_{D-1}$ is a hypersurface in $M_{D}$ given by
\bq
\lb{B.1}
M_{D-1} = \left\{x^{a}: \Phi\left(x^{c}\right) = 0\right\}.
\eq
If we choose the intrinsic coordinates of $M_{D-1}$ as
\bq
\lb{B.2}
\left\{\xi^{\mu}\right\} = \left\{\xi^{0}, \xi^{2}, ..., \xi^{D-2}\right\},
\eq
we find that the hypersurface $M_{D-1}$ can be also written in the form,
\bq
\lb{B.3}
x^{a} = x^{a}(\xi^{\mu}).
\eq
Then, we have
\bq
\lb{B.4}
d\Phi(x^{c}) = \frac{\partial \Phi(x^{c})}{\partial x^{a}}
\frac{\partial x^{a}(\xi^{\nu})}{\partial \xi^{\lambda}}d\xi^{\lambda} = 0.
\eq
Since $d\xi^{\lambda}$'s are linearly independent, we must have
\bq
\lb{B.5}
N_{a} e^{a}_{(\mu)} = 0,
\eq
where
\bqn
\lb{B.6}
N_{a} &\equiv& \frac{\partial \Phi(x^{c})}{\partial x^{a}},\nb\\
e^{a}_{(\mu)}&\equiv& \frac{\partial x^{a}(\xi^{\nu})}{\partial 
\xi^{\mu}},
\eqn
and $N_{a}$ denotes the normal vector to the hypersurface $\Phi(x^{c}) = 0$,
and $e^{a}_{(\mu)}$'s are the tangent vectors. 

When $N_{a}N^{a} \not= 0$, a condition that we shall assume in this section, 
we define the unit normal vector $n_{a}$ as
\bq
\lb{B.7}
n_{a} = \frac{N_{a}}{\left|N_{c}N^{c}\right|^{1/2}},
\eq
with
\bq
\lb{B.8}
n_{a} n_{b} g^{ab} = \epsilon(n),
\eq
where $\epsilon(n) = \pm 1$. 
When $\epsilon(n) = + 1$ the normal vector $n_{a}$ is timelike, and the 
corresponding hypersurface $M_{D-1}$ is spacelike; when $\epsilon(n) = - 1$ 
the normal vector $n_{a}$ is spacelike, and the corresponding hypersurface 
$M_{D-1}$ is timelike.

On the hypersurface $M_{D-1}$, the metric (\ref{A.1}) reduces to
\bqn
\lb{B.9}
\left. ds^{2}\right|_{M_{D-1}} &=&  
g_{ab}\left(x^{c}(\xi^{\lambda})\right)\frac{\partial 
x^{a}(\xi^{\rho})}{\partial \xi^{\mu}} \frac{\partial 
x^{b}(\xi^{\sigma})}{\partial \xi^{\nu}} d\xi^{\mu}d\xi^{\nu}\nb\\
&=& g_{\mu\nu}(\xi^{\lambda}) d\xi^{\mu}d\xi^{\nu},
\eqn
where $g_{\mu\nu}$ is {\em the reduced metric} on $M_{D-1}$ and defined as
\bq
\lb{B.10}
g_{\mu\nu}(\xi^{\lambda}) \equiv 
g_{ab}\left(x^{c}(\xi^{\lambda})\right)e^{a}_{(\mu)}e^{b}_{(\nu)}.
\eq

On the other hand, introducing the projection operator, $h_{ab}$, by
\bq
\lb{B.11}
h_{ab} = g_{ab} - \epsilon(n) n_{a}n_{b},
\eq
we find the following useful relations,
\bqn
\lb{B.12}
g^{ab} &=& g^{\mu\nu}e_{(\mu)}^{a}e_{(\nu)}^{b} + \epsilon(n) 
n^{a}n^{b},\nb\\
g_{\mu\nu} &=& g_{ab} e^{a}_{(\mu)}e^{b}_{(\nu)},\nb\\
h_{ab} &= & g_{ab} - \epsilon(n) n_{a}n_{b} \nb\\
&=&  g^{\mu\nu}e_{(\mu)\;a}e_{(\nu)\; b},
\eqn
where $e_{(\mu)\; a} \equiv g_{ab}e_{(\mu)}^{b}$.

For a tangent vector $\bf{A}$ of $M_{D-1}$, we have
\bq
\lb{B.14}
A_{\mu} = {\bf e}_{(\mu)}\cdot {\bf A} = e^{c}_{(\mu)} A_{c},\;\;\; 
{\bf{A}} = A^{\mu}{\bf e}_{(\mu)},
\eq
with ${\bf A} \cdot {\bf n} = 0$, and 
\bq
\lb{B.14a}
A^{\mu} \equiv g^{\mu\nu}A_{\nu}.
\eq
The intrinsic covariant derivative of ${\bf A}$ with respect to $\xi^{\mu}$ is 
defined as the projection of the vector $\nabla {\bf A}/\nabla\xi^{\mu}$ onto $M_{D-1}$,
\bqn
\lb{B.15}
A_{\mu;\nu} &\equiv& {\bf e}_{(\mu)}\cdot \frac{\nabla {\bf A}}{\nabla\xi^{\nu}}
= e^{c}_{(\mu)}\frac{\partial x^{b}}{\partial \xi^{\nu}}\nabla_{b}A_{c}\nb\\
&=& \frac{\partial x^{b}}{\partial\xi^{\nu}}
\left[\nabla_{b}\left(e^{c}_{(\mu)}A_{c}\right) - 
A_{c}\nabla_{b}\left(e^{c}_{(\mu)}\right)\right]\nb\\
&=&\frac{\partial x^{b}}{\partial\xi^{\nu}} 
\nabla_{b}\left(e^{c}_{(\mu)}A_{c}\right)
- {\bf A} \cdot \frac{\nabla}{\nabla \xi^{\nu}}\left({\bf e}_{(\mu)}\right).\nb\\
\eqn
Since
\bqn
\lb{B.16}
\frac{\nabla}{\nabla \xi^{\nu}}\left({\bf e}_{(\mu)}\cdot {\bf A}\right)
&=& \frac{\partial x^{c}}{\partial \xi^{\nu}}\nabla_{c}\left(A_{\mu}\right) = 
\frac{\partial A_{\mu}}{\partial \xi^{\nu}},\nb\\
{\bf A} \cdot \frac{\nabla}{\nabla \xi^{\nu}}\left({\bf e}_{(\mu)}\right) 
&=& 
A^{\sigma}{\bf e}_{(\sigma)}\cdot\frac{\nabla}{\nabla\xi^{\nu}}\left({\bf 
e}_{(\mu)}\right),
\eqn
we find that Eq.(\ref{B.15}) can be written as
\bq
\lb{B.17}
A_{\mu;\nu} = {\bf e}_{(\mu)}\cdot \frac{\nabla {\bf A}}{\nabla\xi^{\nu}}
= A_{\mu,\nu} - A_{\lambda}\Gamma^{\lambda}_{\mu\nu}
\eq
where
\bq
\lb{B.18}
\Gamma^{\lambda}_{\mu\nu} \equiv g^{\lambda\sigma}{\bf e}_{(\sigma)} \cdot 
\frac{\nabla {\bf e}_{(\mu)}}{\nabla\xi^{\nu}}.
\eq
After  tedious but simple calculations, we finally arrive at
\bqn
\lb{B.19}
\Gamma^{\lambda}_{\mu\nu} &\equiv& g^{\lambda\sigma}{\bf e}_{(\sigma)} \cdot 
\frac{\nabla {\bf e}_{(\mu)}}{\nabla\xi^{\nu}}\nb\\
&=&\frac{1}{2}g^{\lambda\sigma}\left(g_{\sigma \nu,\mu} + g_{\mu\sigma, \nu}
- g_{\mu\nu,\sigma}\right).
\eqn

Properties of a non-intrinsic character enter when we consider the way in 
which $M_{D-1}$ bends in $M_{D}$. This is measured by the variations of $\nabla 
n_{A}/\nabla \xi^{\mu}$ of the normal vector. Since each of these $(D-1)$ 
vectors is perpendicular to $n_{a}$, we can write
\bq
\lb{B.20}
\frac{\nabla n^{a}}{\nabla\xi^{\nu}} = K^{\lambda}_{\nu} e^{a}_{(\lambda)},
\eq
thus defining the extrinsic curvature $K_{\mu\nu}$ of the hypersurface 
$M_{D-1}$. From Eqs.(\ref{B.12}) and (\ref{B.20}) we obtain that
\bqn
\lb{B.21}
K_{\mu\nu} &=& g_{\mu\lambda} K^{\lambda}_{\nu} = e_{(\mu)\; a} 
e^{a}_{(\lambda)} K^{\lambda}_{\nu} \nb\\ 
&=& e_{(\mu)\; a} \frac{\nabla n^{a}}{\nabla\xi^{\nu}}
= e^{a}_{(\mu)} e^{b}_{(\nu)}\nabla_{b}n_{a}.
\eqn
Because   $n_{a} e^{a}_{(\mu)} = 0$,  we find that
\bqn
\lb{B.22}
K_{\mu\nu} &=& e^{a}_{(\mu)}e^{b}_{(\nu)}\nabla_{b}n_{a} 
= - n_{a}e^{b}_{(\nu)}\nabla_{b}\left(e^{a}_{(\mu)}\right)\nb\\
&=& - n_{a}e^{b}_{(\nu)}\left(e^{a}_{(\mu),b}
+ {}^{(D)}\Gamma^{a}_{bc}e^{c}_{(\mu)}\right)\nb\\
&=& - n_{a}\left(\frac{\partial^{2} x^{a}}{\partial \xi^{\mu} 
\partial \xi^{\nu}} + {}^{(D)}\Gamma^{a}_{bc}
\frac{\partial x^{b}}{\partial \xi^{\nu}}
\frac{\partial x^{c}}{\partial \xi^{\mu}}\right)\nb\\
&=& K_{\nu\mu}.
\eqn

Assuming
\bq
\lb{B.23}
\frac{\nabla{\bf e}_{(\mu)}}{\nabla\xi^{\nu}} = \alpha_{\mu\nu}{\bf n}
+ \beta^{\sigma}_{\mu\nu}{\bf e}_{(\sigma)},
\eq
we find that
\bqn
\lb{B.24}
{\bf n} \cdot \frac{\nabla{\bf e}_{(\mu)}}{\nabla\xi^{\nu}} &=& 
\alpha_{\mu\nu} \epsilon(n) = - K_{\mu\nu},\nb\\
{\bf e}_{(\lambda)} \cdot \frac{\nabla{\bf e}_{(\mu)}}{\nabla\xi^{\nu}} &=&
\beta^{\sigma}_{\mu\nu} g_{\lambda\sigma} 
= g_{\lambda\sigma}\Gamma^{\sigma}_{ \mu\nu},
\eqn
namely,
\bq
\lb{B.25}
\alpha_{\mu\nu} = - \epsilon(n) K_{\mu\nu},\;\;\;\;
\beta^{\sigma}_{\mu\nu} = \Gamma^{\sigma}_{\mu\nu}.
\eq
Inserting Eq.(\ref{B.25}) into Eq.(\ref{B.23}), we obtain
\bq
\lb{B.26}
\frac{\nabla{\bf e}_{(\mu)}}{\nabla\xi^{\nu}} = 
-  \epsilon(n) K_{\mu\nu}{\bf n}
+ \Gamma^{\sigma}_{\mu\nu}{\bf e}_{(\sigma)},
\eq
which is usually called the {\em Gauss-Weingarten} equation. Thus, for any 
vector ${\bf A}$ that is tangent to $M_{D-1}$, we have
\bqn
\lb{B.27}
\frac{\nabla{\bf A}}{\nabla\xi^{\nu}} &=& \frac{\nabla}{\nabla\xi^{\nu}}
\left(A^{\mu}{\bf e}_{(\mu)}\right) \nb\\
&=& \frac{\nabla A^{\mu}}{\nabla\xi^{\nu}}{\bf e}_{(\mu)} + A^{\mu} \frac{\nabla {\bf 
e}_{(\mu)}}{\nabla\xi^{\nu}} \nb\\
&=& \frac{\partial A^{\mu}}{\partial \xi^{\nu}}
{\bf e}_{(\mu)} + A^{\mu}\left(-\epsilon(n) K_{\mu\nu} {\bf n}
+ \Gamma^{\sigma}_{\mu\nu}{\bf e}_{(\sigma)}\right)\nb\\
&=& A^{\mu}_{\;\; ; \nu} {\bf e}_{(\mu)} - \epsilon(n) A^{\mu}K_{\mu\nu} {\bf 
n},\nb
\eqn
that is,
\bq
\lb{B.27a}
\frac{\nabla{\bf A}}{\nabla\xi^{\nu}} = A^{\mu}_{\;\; ; \nu} {\bf e}_{(\mu)} - 
\epsilon(n) A^{\mu}K_{\mu\nu} {\bf n}.
\eq

Operating on Eq.(\ref{B.26}) with $\nabla/\nabla\xi^{\lambda}$ and using 
Eq.(\ref{B.20}), we find that
\bqn
\lb{B.28}
  \frac{\nabla}{\nabla\xi^{\lambda}}\left(\frac{\nabla e^{a}_{(\mu)}}{\nabla\xi^{\nu}}\right)
&=& \frac{\nabla}{\nabla\xi^{\lambda}}\left(-\epsilon(n)K_{\mu\nu}n^{a} + 
\Gamma^{\sigma}_{\mu\nu}e^{a}_{(\sigma)}\right)\nb\\
&=& - \epsilon(n)\frac{\nabla K_{\mu\nu}}{\nabla\xi^{\lambda}}n^{a}
- \epsilon(n) K_{\mu\nu}\frac{\nabla n^{a}}{\nabla\xi^{\lambda}}\nb\\
& &
+ \frac{\nabla \Gamma^{\sigma}_{\mu\nu}}{\nabla\xi^{\lambda}}e^{a}_{(\sigma)}
+ \Gamma^{\delta}_{\mu\nu}\frac{\nabla e^{a}_{(\delta)}}{\nabla\xi^{\lambda}}\nb\\
&=&  - \epsilon(n) K_{\mu\nu,\lambda} n^{a} - 
\epsilon(n)K_{\mu\nu}K^{\sigma}_{\lambda} e^{a}_{(\sigma)} \nb\\
& & 
+ \Gamma^{\sigma}_{\mu\nu,\lambda} e^{a}_{(\sigma)} \nb\\
& &
 + \Gamma^{\delta}_{\mu\nu}\left(- \epsilon(n) 
K_{\delta\lambda}n^{a} 
+ \Gamma^{\sigma}_{\delta\lambda}e^{a}_{(\sigma)}\right)\nb\\
&=&  \left(\Gamma^{\sigma}_{\mu\nu,\lambda} + \Gamma^{\delta}_{\mu\nu}
\Gamma^{\sigma}_{\delta\lambda} - 
\epsilon(n)K_{\mu\nu}K^{\sigma}_{\lambda}\right) e^{a}_{(\sigma)} \nb\\
& &
 - \epsilon(n)\left(K_{\mu\nu,\lambda} + 
\Gamma^{\delta}_{\mu\nu}K_{\delta\lambda}\right)n^{a}.
\eqn 
Thus, we have
\bqn
\lb{B.29}
  \left(\frac{\nabla^{2}}{\nabla\xi^{\lambda}\nabla\xi^{\nu}} \right.
&-& \left. \frac{\nabla^{2}}{\nabla\xi^{\nu} \nabla\xi^{\lambda}}\right)e^{a}_{(\mu)} =
{}^{(D-1)}R^{\sigma}_{\mu\lambda\nu} e^{a}_{(\sigma)} \nb\\
& &
 + \epsilon(n)\left(K_{\mu\lambda}K^{\sigma}_{\nu} - 
K_{\mu\nu}K^{\sigma}_{\lambda}\right)e^{a}_{(\sigma)}\nb\\
& &
+ \epsilon(n)\left(K_{\mu\lambda;\nu}-
K_{\mu\nu;\lambda}\right) n^{a},
\eqn
where
\bq
\lb{B.30}
 {}^{(D-1)}R^{\sigma}_{\mu\lambda\nu} \equiv
\Gamma^{\sigma}_{\mu\nu,\lambda} 
- \Gamma^{\sigma}_{\mu\lambda,\nu}
+ \Gamma^{\delta}_{\mu\nu}\Gamma^{\sigma}_{\delta\lambda}
- \Gamma^{\delta}_{\mu\lambda}\Gamma^{\sigma}_{\delta\nu}.
\eq
On the other hand, we have
\bqn
\lb{B.31a}
 \frac{\nabla^{2}e^{a}_{(\mu)}}{\nabla\xi^{\lambda}\nabla\xi^{\nu}} &=& 
\frac{\nabla}{\nabla\xi^{\lambda}}
\left(\frac{\nabla e^{a}_{(\mu)}}{\nabla\xi^{\nu}}\right) 
= e^{c}_{(\lambda)}\nabla_{c}\left(e^{b}_{(\nu)}\nabla_{b}e^{a}_{(\mu)}\right)\nb\\
&=&  e^{c}_{(\lambda)}e^{b}_{(\nu)} \left(\nabla_{c} \nabla_{b}e^{a}_{(\mu)}\right)\nb\\
& & 
+ e^{c}_{(\lambda)}\left(\nabla_{c}e^{b}_{(\nu)}\right) 
\left(\nabla_{b}e^{a}_{(\mu)}\right)\nb\\
&=&  e^{c}_{(\lambda)}e^{b}_{(\nu)} \left(\nabla_{c} \nabla_{b}e^{a}_{(\mu)}\right)
+ \left(\nabla_{b}e^{a}_{(\mu)}\right)\nb\\
& &
\times\left(\frac{\partial^{2} x^{b}}{\partial 
\xi^{\lambda}\partial\xi^{\nu}} + {}^{(D)}\Gamma^{b}_{cd}
\frac{\partial x^{c}}{\partial\xi^{\lambda}}
\frac{\partial x^{d}}{\partial\xi^{\nu}}\right),
\eqn
and
\bqn
\lb{B.31b}
\left(\frac{\nabla^{2}}{\nabla\xi^{\lambda}\nabla\xi^{\nu}}\right.
&-&\left. \frac{\nabla^{2}}{\nabla\xi^{\nu}\nabla\xi^{\lambda}}\right) e^{a}_{(\mu)}\nb\\
&=& \left[\left(\nabla_{c}\nabla_{b} - \nabla_{b}\nabla_{c}\right)e^{a}_{(\mu)}\right]
e^{c}_{(\lambda)}e^{b}_{(\nu)}\nb\\
&=&   {}^{(D)}R^{a}_{dcb}e^{d}_{(\mu)}e^{c}_{(\lambda)}e^{b}_{(\nu)}.
\eqn
Then, the combination of Eqs.(\ref{B.29}) and (\ref{B.31b}) yields,
\bqn
\lb{B.32}
 {}^{(D)}R^{ a}_{dcb}e^{d}_{(\mu)}e^{c}_{(\lambda)}e^{b}_{(\nu)}  &=&
 {}^{(D-1)}R^{\sigma}_{\mu\lambda\nu} e^{a}_{(\sigma)} \nb\\
& &
 + \epsilon(n)\left(K_{\mu\lambda}K^{\sigma}_{\nu} - 
K_{\mu\nu}K^{\sigma}_{\lambda}\right)e^{a}_{(\sigma)}\nb\\
& & + \epsilon(n)\left(K_{\mu\lambda;\nu}
- K_{\mu\nu;\lambda}\right) n^{a}.\nb\\
\eqn
Multiplying Eq.(\ref{B.32})  by $e_{(\rho) \; a}$  
we obtain the {\em Gauss   equation},
\bqn
\lb{B.33a} 
& &  R^{(D)}_{abcd} e^{a}_{(\rho)} e^{b}_{(\mu)}e^{c}_{(\lambda)}e^{d}_{(\nu)}
=  R^{(D-1)}_{\rho\mu\lambda\nu}  \nb\\
&&\;\;\;\;\; + \epsilon(n)\left( K_{\mu\lambda}K_{\nu\rho} - 
K_{\mu\nu}K_{\lambda\rho}\right).
\eqn
Similarly, multiplying Eq.(\ref{B.32}) with $n_{a}$
we obtain the {\em Codacci equation},
\bq
\lb{B.33b} 
 R^{(D)}_{abcd} n^{a} e^{b}_{(\mu)}e^{c}_{(\lambda)}e^{d}_{(\nu)}
=  K_{\mu\lambda;\nu} - K_{\mu\nu;\lambda}.
\eq
Multiplying Eq.(\ref{B.33a}) by $g^{\rho\lambda}g^{\mu\nu}$, and noting
\bq
\lb{B.34}
g^{\mu\nu}e^{a}_{(\mu)}e^{b}_{(\nu)} = g^{ab} - \epsilon(n)n^{a}n^{b},
\eq
we find that
\bqn
\lb{B.35}
& &   R^{(D)}_{abcd} e^{a}_{(\rho)} e^{b}_{(\mu)}e^{c}_{(\lambda)}e^{d}_{(\nu)}
g^{\rho\lambda}g^{\mu\nu}  
=  R^{(D)}_{abcd}\nb\\
& & \;\;\;\;\; \times \left(g^{ac} - 
\epsilon(n)n^{a}n^{c}\right)\left(g^{bd} - 
\epsilon(n)n^{b}n^{d}\right)\nb\\
& &  =  R^{(D)}_{abcd} \left(g^{ac}g^{bd} 
- \epsilon(n) g^{ac}n^{b}n^{d}\right.\nb\\
& & \;\;\;\;\;\left.
- \epsilon(n) g^{bd}n^{a}n^{c}\right)\nb\\
& & =  R^{(D)} - 2\epsilon(n) \;  R^{(D)}_{ab}n^{a}n^{b}\nb\\
& & = - 2\epsilon(n) \;  G^{(D)}_{ab}n^{a}n^{b}\nb\\
& & =  R^{(D-1)} + \epsilon(n)\left(K^{\lambda}_{\sigma}K_{\lambda}^{\sigma}
- K^{2}\right),\nb
\eqn
this is,
\bq
\lb{B.36}
 - 2\epsilon(n) \;  G^{(D)}_{ab}n^{a}n^{b}
 = R^{(D-1)} + \epsilon(n)\left(K^{\lambda}_{\sigma}K_{\lambda}^{\sigma}
- K^{2}\right),
\eq
where $K = g^{\mu\nu}K_{\mu\nu}$.

Multiplying Eq.(\ref{B.33b}) by $g^{\mu\nu}$, we obtain
\bqn
\lb{B.37}
 R^{(D)}_{abcd} n^{a} e^{b}_{(\mu)}e^{c}_{(\lambda)}e^{d}_{(\nu)} g^{\mu\nu}
&=&  R^{(D)}_{abcd} n^{a}  e^{c}_{(\lambda)}\nb\\
& & \times \left(g^{bd} 
- \epsilon(n) n^{b}n^{d}\right)\nb\\
&=&  R^{(D)}_{ac}n^{a}e^{c}_{(\lambda)}\nb\\
&=&  G^{(D)}_{ac}n^{a}e^{c}_{(\lambda)}\nb\\
&=& \left(K^{\sigma}_{\lambda} - \delta^{\sigma}_{\lambda} K\right)_{;\sigma},\nb
\eqn
or
\bq
\lb{B.38}
 G^{(D)}_{ac}n^{a}e^{c}_{(\lambda)}
= \left(K^{\sigma}_{\lambda} - \delta^{\sigma}_{\lambda} K\right)_{;\sigma}.
\eq

From the Gauss equation Eq.(\ref{B.33a}),   we find that
\bqn
\lb{C.1} 
 R^{(D-1)}_{\rho\mu\lambda\nu} &=&  R^{(D)}_{abcd} e^{a}_{(\rho)} 
e^{b}_{(\mu)}e^{c}_{(\lambda)}e^{d}_{(\nu)} \nb\\
& & 
 - \epsilon(n)\left( K_{\mu\lambda}K_{\nu\rho} - 
K_{\mu\nu}K_{\lambda\rho}\right),
\eqn
from which we obtain
\bqn
\lb{C.2}
 R^{(D-1)}_{\mu\nu} &=&  R^{(D)}_{ab}e^{a}_{(\mu)}e^{b}_{(\nu)}\nb\\
& & 
- \epsilon(n) \;  R^{(D)}_{abcd}n^{a}e^{b}_{(\mu)}n^{c}e^{d}_{(\nu)} \nb\\
& &
- \epsilon(n)\left(K_{\mu\sigma}K^{\sigma}_{\nu} - K 
K_{\mu\nu}\right),\nb\\
R^{(D-1)} &=&  R^{(D)} - 2\epsilon(n) \; R^{(D)}_{ab}n^{a} n^{b} \nb\\
& &
 - \epsilon(n)\left(K_{\alpha\beta}K^{\alpha\beta} - K^{2}\right).
\eqn
Then, from Eq.(\ref{A.8}) we find that
\bqn
\lb{C.3}
& &  R^{(D)}_{abcd}n^{a}e^{b}_{(\mu)}n^{c}e^{d}_{(\nu)}  
= \frac{\epsilon(n)}{D-2}\;  R^{(D)}_{ab}e^{a}_{(\mu)}e^{b}_{(\nu)}\nb\\
& & + \frac{1}{D-2}\left\{R^{(D)}_{ab}n^{a}n^{b} 
- \frac{\epsilon(n)}{D-1}\;  R^{(D)}\right\}g_{\mu\nu}\nb\\
& & + \;  E^{(D)}_{\mu\nu},
\eqn
where
\bq
\lb{C.4}
E^{(D)}_{\mu\nu} \equiv \; 
 C^{(D)}_{abcd}n^{a}e^{b}_{(\mu)}n^{c}e^{d}_{(\nu)}.
\eq

From the definition of the Einstein tensor, we  find that
\bqn
\lb{C.5}
R^{(D)}_{ab}e^{a}_{(\mu)}e^{b}_{(\nu)} &=&  
  G^{(D)}_{ab} e^{a}_{(\mu)}e^{b}_{(\nu)}
- \frac{1}{D-2}g_{\mu\nu}\;  G^{(D)},\nb\\
 R^{(D)}_{ab}n^{a}n^{b}  &=&  
   G^{(D)}_{ab} n^{a}n^{b}
- \frac{\epsilon(n)}{D-2}   G^{(D)},\nb\\
 R^{(D)} &=& -  \frac{2}{D-2}\;  G^{(D)}.
\eqn
Then, combining Eqs.(\ref{C.2})-(\ref{C.5}), we obtain
\bqn
\lb{C.6}
 G^{(D-1)}_{\mu\nu} &=&  \frac{D-3}{D-2}
\left\{G^{(D)}_{ab}e^{a}_{(\mu)} e^{b}_{(\nu)} \right.\nb\\
& & + \epsilon(n)  G^{(D)}_{ab}n^{a}n^{b}g_{\mu\nu} \nb\\
& & \left.
- \frac{1}{D-1} G^{(D)} g_{\mu\nu}\right\}\nb\\
& & - \epsilon(n)\left(K_{\mu\sigma}K^{\sigma}_{\nu} - K K_{\mu\nu}\right)\nb\\
& & 
+ \frac{\epsilon(n)}{2}\left(K_{\alpha\beta}K^{\alpha\beta} - 
K^{2}\right)g_{\mu\nu} \nb\\
& & 
 - \epsilon(n) \;  E^{(D)}_{\mu\nu}.
\eqn

\subsection{Surface Layers}

\renewcommand{\theequation}{C.\arabic{equation}}
\setcounter{equation}{0}

Assume that the hypersurface $M_{D-1}$ divides the whole spacetime $M_{D}$ into two 
regions $M^{\pm}_{D}$, where 
\bq
\lb{D.1}
M^{+}_{D} := \left\{x^{+\; a}, \Phi \ge 0\right\},\;\;\;
M^{-}_{D} := \left\{x^{-\; a}, \Phi \le 0\right\}.
\eq
In terms of   $x^{\pm\; a}$, the hypersurface $M_{D-1}$ is 
given by
\bq
\lb{D.2}
x^{+ a} = x^{+a}\left(\xi^{\mu}\right),\;\;\;
x^{- a} = x^{- a}\left(\xi^{\mu}\right),
\eq
or equivalently
\bq
\lb{D.3}
\Phi^{+}\left(x^{+ b}\right) = 0,\;\;\;
\Phi^{-}\left(x^{- b}\right) = 0.
\eq
From the above equations we find that
\bqn
\lb{D.3a}
n^{+}_{a} &=& \frac{N^{+}_{a}}{\left|N^{+}_{c}N^{+c}\right|^{1/2}},\;\;\;
N^{+}_{a} = \frac{\partial \Phi^{+}\left(x^{+c}\right)}{\partial x^{+a}},\nb\\
e^{+a}_{(\mu)} &\equiv& \frac{\partial x^{+ a}\left(\xi^{\lambda}\right)}
    {\partial \xi^{\mu}}, \nb\\
n^{-}_{a} &=& \frac{N^{-}_{a}}{\left|N^{-}_{c}N^{-c}\right|^{1/2}},\;\;\;
N^{-}_{a} = \frac{\partial \Phi^{-}\left(x^{- c}\right)}
         {\partial x^{- a}},\nb\\
e^{- a}_{(\mu)} &\equiv& \frac{\partial x^{- a}\left(\xi^{\lambda}\right)}
          {\partial \xi^{\mu}}.  
\eqn
Then, it is easy to see that in each of the two regions, the Gauss and Codacci 
equations take the form of Eqs.(\ref{B.33a}) and (\ref{B.33b}), from which 
Eqs.(\ref{B.36}) and (\ref{B.38}) result. 
On the hypersurface $M_{D-1}$, the reduced metric from each side of $M_{D-1}$ 
should be the same, so we must have
\bq
\lb{D.4}
\left. g^{+}_{\mu\nu}\left(\xi^{\mu}\right)\right|_{\Sigma^{+}} 
= \left. g^{-}_{\mu\nu}\left(\xi^{\mu}\right)\right|_{\Sigma^{-}}
\equiv g_{\mu\nu}\left(\xi^{\mu}\right).
\eq
On the other hand, from the Lanczos equations \cite{Lan22},
\bq
\lb{D.5}
\left[K_{\mu\nu}\right]^{-} - g_{\mu\nu}\left[K\right]^{-} 
= - \kappa^{2}_{D} {\cal{T}}_{\mu\nu},
\eq
one defines the symmetric tensor ${\cal{T}}_{\mu\nu}$ as the effective {\em surface 
energy-momentum tensor}, where
\bqn
\lb{D.6}
\left[K_{\mu\nu}\right]^{-} &\equiv& {\rm lim}_{\Phi \rightarrow 0^{+}}
K^{+}_{\mu\nu} - {\rm lim}_{\Phi \rightarrow 0^{-}}
K^{-}_{\mu\nu},\nb\\
\left[K\right]^{-} &\equiv& g^{\mu\nu}\left[K_{\mu\nu}\right]^{-}.
\eqn
Combining  Eq.(\ref{B.38}) with Eq.(\ref{D.5}), we obtain that
\bq
\lb{D.8}
\left[G^{(D)}_{ac}n^{a}e^{c}_{(\mu)}\right]^{-} = - \kappa_{D} 
{\cal{T}}^{\lambda}_{\mu\; ; 
\lambda},
\eq
which serves as the conservation law for the surface EMT.

Assuming reflection symmetry of the brane, we have
\bq
\lb{D.11}
K^{+}_{\mu\nu} = - K^{-}_{\mu\nu} = - K_{\mu\nu}.
\eq
Then, from the Lanczos equations (\ref{D.5}) we find
that
\bq
\lb{D.12}
K_{\mu\nu} - g_{\mu\nu} K =  \frac{\kappa^{2}_{D}}{2}{\cal{T}}_{\mu\nu}.
\eq
Considering the case where
\bq
\lb{D.13}
{\cal{T}}_{\mu\nu} = \tau_{\mu\nu} + \lambda^{total} g_{\mu\nu},
\eq
we find that 
\bqn
\lb{D.14}
K &=& - \frac{\kappa^{2}_{D}}{2(D-2)}\left[(D-1)\lambda^{total} + \tau\right],\nb\\
K_{\mu\nu} &=& \frac{\kappa^{2}_{D}}{2}\left[\tau_{\mu\nu}
- \frac{1}{D-2}\left(\tau + \lambda^{total}\right)g_{\mu\nu}\right],
\eqn
where $\tau \equiv g^{\mu\nu}\tau_{\mu\nu}$, and 
\bq
\lb{D.14a}
\lambda^{total} \equiv \lambda + \tau_{p}. 
\eq
Then, we obtain
\bqn
\lb{D.16}
{\cal{F}}^{(D-1)}_{\mu\nu} &\equiv& \left(K_{\mu\lambda}K^{\lambda}_{\nu} - KK_{\mu\nu}\right) \nb\\
& &
   - \frac{1}{2}g_{\mu\nu}\left(K_{\alpha\beta}K^{\alpha\beta} -
   K^{2}\right)\nb\\
   &=& -\epsilon(n)\left\{\kappa^{2}_{D-1}\tau_{\mu\nu} + \lambda^{eff.} g_{\mu\nu} 
   + \kappa^{4}_{D}\pi_{\mu\nu}\right\}\nb\\
   & & + \frac{\kappa^{4}_{D}(D-3)}{4(D-2)}\tau_{p}
   \left\{\tau_{\mu\nu} 
   + \frac{1}{2}\left(2\lambda + \tau_{p}\right)g_{\mu\nu}\right\},\nb\\
\eqn
where
\bqn
\lb{D.17}
\pi_{\mu\nu} &=& -\frac{\epsilon(n)}{4}\left\{\tau_{\mu\lambda}\tau^{\lambda}_{\nu}\right.\nb\\
& & \left. 
 -  \frac{1}{D-2}\tau \tau_{\mu\nu}
 - \frac{1}{2}g_{\mu\nu}\left(\tau^{\alpha\beta} \tau_{\alpha\beta}
 - \frac{1}{D-2}\tau^{2}\right)\right\},\nb\\
\kappa^{2}_{D-1} &=& -\epsilon(n)\frac{D-3}{4(D-2)}\lambda\kappa^{4}_{D},\nb\\
\lambda^{eff.} &=& -\epsilon(n)\frac{D-3}{8(D-2)}\lambda^{2}\kappa^{4}_{D}.
\eqn
Then,  Eq.(\ref{C.6}) takes the form,
\bqn
\lb{D.18}
 G^{(D-1)}_{\mu\nu} &=&  - \epsilon(n) \left({\cal{G}}^{(D)}_{\mu\nu}
+  E^{(D)}_{\mu\nu}\right)\nb\\
& & - \epsilon(n) \frac{\kappa^{4}_{D}(D-3)}{4(D-2)}\tau_{p}
   \left\{\tau_{\mu\nu} 
   + \frac{1}{2}\left(2\lambda + \tau_{p}\right)g_{\mu\nu}\right\}\nb\\
& &  + \kappa^{2}_{D-1}\tau_{\mu\nu}
+ \lambda^{eff.} g_{\mu\nu} 
+ \kappa^{4}_{D}\pi_{\mu\nu},
\eqn
where
\bqn
\lb{D.22}
 {\cal{G}}^{(D)}_{\mu\nu} &=& - \epsilon(n)\frac{D-3}{(D-2)}
\left\{G^{(D)}_{ab}e^{a}_{(\mu)} e^{b}_{(\nu)} \right.\nb\\
& & \left.
+ \epsilon(n) \left[G^{(D)}_{ab}n^{a}n^{b} 
- \frac{\epsilon(n)}{D-1} G^{(D)}\right]g_{\mu\nu}\right\}.\nb\\
\eqn

For a perfect fluid,
\bq
\lb{D.19}
\tau_{\mu\nu} = \left(\rho + p\right)u_{\mu}u_{\nu} - p g_{\mu\nu},
\eq
where $u_{\mu}$ is the four-velocity of the fluid, we find that 
\bq
\lb{D.21}
\pi_{\mu\nu} = -\epsilon(n)\frac{D-3}{4(D-2)}
\rho\left\{\left(\rho + p\right)u_{\mu}u_{\nu} 
- \frac{1}{2}(\rho + 2p)g_{\mu\nu}\right\}.
\eq


\begin{thebibliography}{30}
\bibitem{footnote1} It should be noted that our signature is 
different from that used by Israel \cite{Israel66}.
\bibitem{footnote2} The definition for Riemann tensor adopted here is the same as that
used by Israel \cite{Israel66}.
\end{thebibliography}
\end{document}